\documentclass[12pt]{iopart}

\pdfoutput=1

\usepackage{graphicx,sidecap}
\usepackage{bm}
\usepackage{braket}
\usepackage{amssymb}
\usepackage{mathrsfs}
\usepackage{amsmath}
\usepackage{xcolor}
\usepackage{hyperref}
\usepackage{enumerate}
\usepackage[toc,title]{appendix}
\usepackage{verbatim}
\usepackage[margin=30pt, bf, font=footnotesize, center, justification=justified]{caption}[2004/07/16]

\hypersetup{colorlinks,bookmarksopen,bookmarksnumbered,
citecolor=red,
linkcolor=blue,
pdfstartview=green,
urlcolor=purple}

\catcode`@=11 
\renewcommand\tableofcontents{%
  \section*{\contentsname}%
  \@starttoc{toc}%
}
\catcode`@=12

\def\be{\begin{equation}}
\def\ee{\end{equation}}

\def\bea{\begin{eqnarray}}
\def\eea{\end{eqnarray}}

\begin{document}

\title[Entanglement hamiltonians in 2d CFT]
{Entanglement hamiltonians in two-dimensional conformal field theory}

\vspace{.5cm}

\author{John Cardy$^{1,2}$ and Erik Tonni$^3$}
\address{$^1$\,Department of Physics, University of California, Berkeley CA 94720, USA.}
\address{$^2$\,All Souls College, Oxford OX1 4AL, UK.}
\address{$^3$\,SISSA and INFN, via Bonomea 265, 34136 Trieste, Italy.}

\vspace{.5cm}

\begin{abstract}
We enumerate the cases in 2d conformal field theory where the logarithm of the reduced density matrix (the entanglement or modular hamiltonian) may be written as an integral over the energy-momentum tensor times a local weight. These include known examples and new ones corresponding to the time-dependent scenarios of a global and local quench. In these latter cases the entanglement hamiltonian depends on the momentum density as well as the energy density. In all cases the entanglement spectrum is that of the appropriate boundary CFT. We emphasize the role of boundary conditions at the entangling surface and the appearance of boundary entropies as universal $O(1)$ terms in the entanglement entropy.
\end{abstract}

\maketitle

\tableofcontents

\section{Introduction}
\label{sec:intro}

The entanglement hamiltonian, also called the modular hamiltonian, has become an important concept in understanding the nature of entanglement in many-body quantum systems and in quantum field theories. 

Given a bipartition of the Hilbert space ${\cal H}={\cal H}_A\otimes{\cal H}_B$, and a density matrix $\rho$, one defines the entanglement hamiltonian of the subsystem $A$ by\footnote{It is convenient to insert a factor of $2\pi$ into this definition, see Eq.~(\ref{halfspace}).}
\be
K_A=-\frac1{2\pi}\log\rho_A\,,
\ee
where $\rho_A={\rm Tr}_{{\cal H}_B}\,\rho$ is the reduced density matrix of the subsystem. 

The knowledge of the spectrum of $K_A$ determines all the R\'enyi entropies 
\be
S_A^{(n)}=-\frac1{n-1}\log{\rm Tr}_{{\cal H}_A}e^{-2\pi nK_A}\,,
\ee
and the entanglement entropy $S_A=\lim_{n\to1}S^{(n)}_A=2\pi\,{\rm Tr}_{{\cal H}_A}K_Ae^{-2\pi K_A} $.

The entanglement, or modular, hamiltonian plays a central role in computing relative entropies and the so-called first law of entanglement \cite{blanco-13}.
The entanglement spectrum has been useful for understanding the nature of entanglement in gapped states in 2+1-dimensional quantum many-body systems \cite{haldane-08}, in 1+1-dimensional integrable models \cite{peschel-04}, and in relativistic quantum field theories, in particular conformal field theories (CFTs). In relativistic field theories there are special cases when $K_A$ may be expressed as an integral over the local energy-momentum tensor with a suitable weight factor. An important example is when the theory is defined on the whole of flat Minkowski space (or $\mathbb{R}^d$ in euclidean space), $\rho$ corresponds to the vacuum state, and $A$ consists of the degrees of freedom in a half-space $x_1>0$, in which case\footnote{Note that here we do \em not \em include the factor of $1/2\pi$ multiplying the energy-momentum tensor which is conventional in 2d CFT.} \cite{bw, unruh}
\be\label{halfspace}
K_A=\int_Ax_1 \,T_{00}(x)d^{d-1}x\,,
\ee 
\em i.e. \em a generator of Lorentz boosts, or, equivalently, of euclidean rotations. 

For a CFT this result may be conformally mapped into the case when $A$ is a ball of radius $R$, giving \cite{chm, longo}
\be\label{KAball}
K_A=\int_A\frac{R^2-x^2}{2R} \;T_{00}(x)d^{d-1}x\,.
\ee
In 1+1 dimensions, $A$ is a single interval of length $2R$ in an infinite system. 

In this paper we enumerate other cases in 2d CFT when $K_A$ may be written as a local integral over the energy-momentum tensor. These include known examples such as the above case of a single interval, and its generalizations to finite size and finite temperature (but not both) \cite{borchers-98, klich-13}, but also new time-dependent scenarios when $\rho$ corresponds to a pure state following a global or local quantum quench of the kind first considered in Refs.~\cite{cc-05-global quench,cc-07-local quench}. 

\begin{figure}
\vspace{.2cm}
\hspace{1.55cm}
\includegraphics[width=0.8\textwidth]{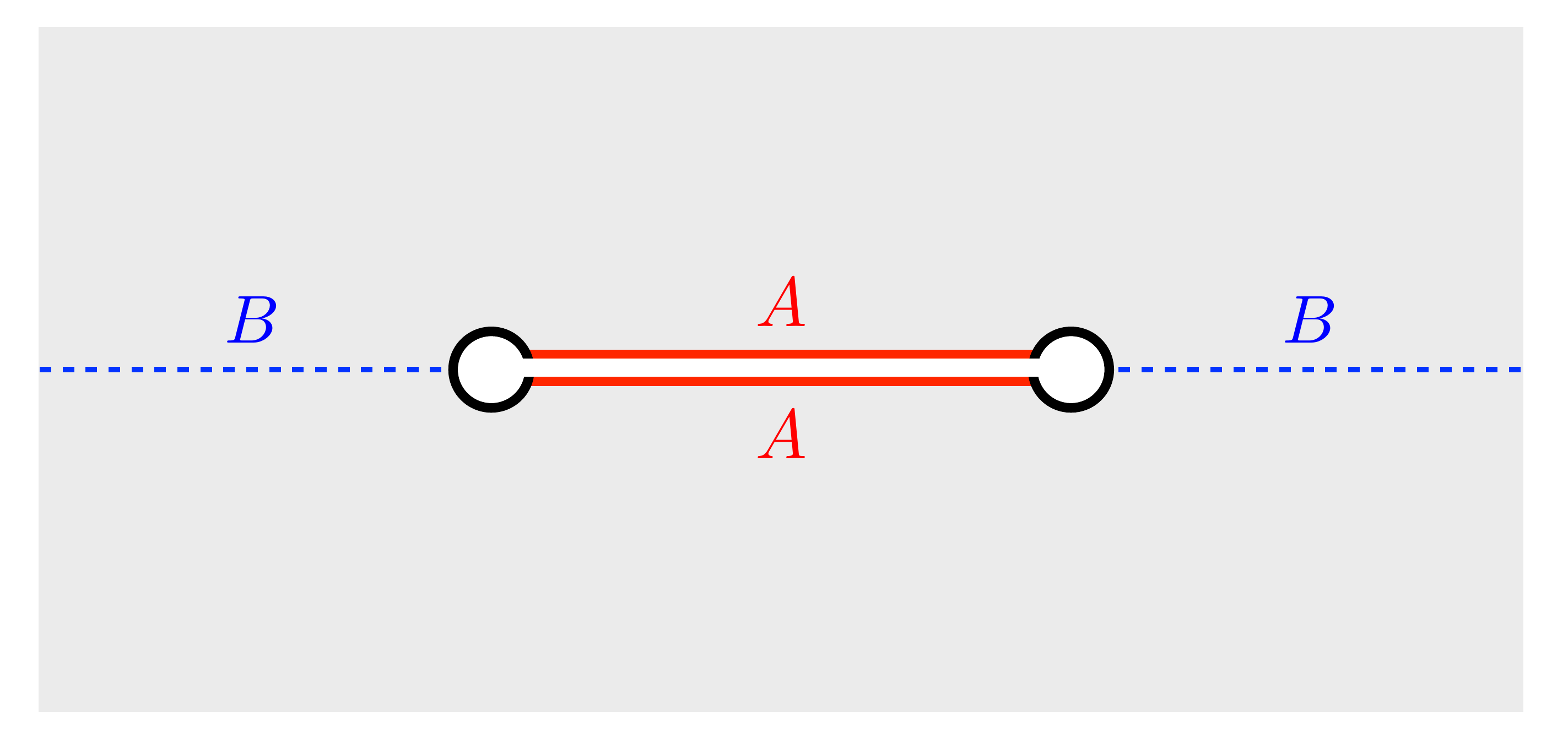}
\vspace{.1cm}
\caption{
Euclidean space-time region for the path integral for reduced density matrix $\rho_A$ of the projected state $P_{\partial A}^\epsilon|\psi\rangle$. As usual the rows and columns of the density matrix are labelled by the values of the fields on the upper and lower edges of the slit along $A$ (shown in red.) The projection induces a boundary condition on the parts of the slit within $\epsilon $ of the boundary points between $A$ and $B$, shown in black. When the moments ${\rm Tr}\,\rho_A^n$ are computed, $n$ copies of this picture are sewn together cyclically along the red edges, but this leaves small black-edged holes around the boundary points. When $A$ is a single interval in an infinite system, the resulting manifold is topologically an annulus.}
\label{fig:PAB}
\end{figure}

All the examples we consider have one feature in common. First, it is necessary to recognize that for a quantum field theory the expressions for the R\'enyi entropies and the entanglement entropy are UV divergent and must be regularized in some way. The underlying reason for this is that the above sharp bipartition of the Hilbert space into subspaces corresponding to sharp spatial regions, and the associated division of the operator algebras, are not legitimate, since local operators are distribution-valued and must be smeared spatially against test functions. This difficulty does not appear on the lattice (at least for theories without local gauge invariance), and one simple way \cite{callan-94, holzhey-94} to bypass it for a field theory is to consider only states in the Hilbert space which are projected onto a common eigenstate of a locally complete set of commuting observables in a small spatial region of thickness $\epsilon$ around the common boundary of $A$ and $B$:
\be
|\psi'\rangle=P_{\partial A}^\epsilon|\psi\rangle\,.
\ee
In the euclidean path integral for the reduced density matrix $\rho_A$, this has the effect of introducing small slits around the entanglement points between $A$ and $B$, and also in the path integral for $S_A^{(n)}$ on the $n$-sheeted cover of this space, see Fig~\ref{fig:PAB}.
This introduces boundaries in the euclidean space-time path integral at each entanglement point between $A$ and $B$, in addition to any other external boundaries which are part of the definition of the state of the full system. The projection onto a common eigenstate of a locally complete set of commuting observables implies a particular choice of boundary conditions on each of these new boundaries. For a CFT, it is natural to choose conformal boundary conditions, corresponding to scale and reparametrization invariance of the boundary state. Although these are special points in the space of all possible boundary conditions, it is believed that they represent stable fixed points of the flows of the boundary renormalization group. Thus any reasonable boundary condition should flow to one of them as long as the scale $\epsilon$ is larger that the UV cut-off scale. Nevertheless, universal features of the results should still be valid if this limit is not necessarily respected, that is, if we take $\epsilon $ to be of the order of the lattice spacing in a lattice theory. Other lattice features should then arise as corrections to IR scaling (although these may be of an `unusual' form \cite{cc-10}.) 

One of the universal features of conformal boundary conditions is the appearance of the Affleck-Ludwig boundary entropy \cite{affleck-ludwig} in the entanglement spectrum \cite{cc-04}. Here we see that this should appear not only from external boundaries but also from the chosen boundary condition in the UV regulator imposed around the entanglement points. They also imply that the spectrum of the entanglement hamiltonian is given in terms of the scaling dimensions of the boundary CFT consistent with the chosen boundary conditions. These observations have already been made by L\"auchli \cite{lauchli} and Ohmori and Tachikawa \cite{tachikawa-14} and have been verified for simple lattice models. 

We are now in position to state a sufficient condition under which $K_A$ may be written as a local integral over the energy-momentum tensor: it holds when the euclidean space-time region, including both external and internal boundaries around the entangling points, is conformally equivalent to an annulus, that is, topologically a sphere with two holes, such that $A$ on a constant time slice is mapped to a simple curve connecting the two boundaries . This includes the above case of a single interval, either in an infinite system in the ground state or at finite temperature, or at zero temperature in finite system with spatially periodic boundary conditions, but not, for example, in a finite system at finite temperature, which would correspond to a torus with two holes. Another example is when $A$ is an interval at the end of a semi-infinite system with one external boundary, and also when 
$A$ is at end of a finite interval $A\cup B$, as long as the boundary conditions at the two ends are the same (when, topologically, they count as a single boundary in euclidean space-time.)

In all these cases we show that the entanglement hamiltonian may  be written as an integral over the local hamiltonian density of the form
\be
\label{KAgeneral}
K_A=\int_A\frac{T_{00}(x)}{f'(x)}dx\,,
\ee
along a constant time slice, where $f(x)$ is the (real) restriction to this time slice of the analytic function $z\to w=f(z)$ which conformally maps the euclidean space-time, with its two boundaries, to an annulus which is a rectangle periodically identified mod$(2\pi)$ in the ${\rm Im}\,w$ direction. In these cases $K_A$ has the same spectrum (above its lowest eigenvalue) as the hamiltonian which generates translations around the annulus in this direction. If $W$ is the width of this annulus in the ${\rm Re}\,w$ direction, these eigenvalues are all of the form  $(\pi/W)(-c/24+\Delta_j)$ \cite{jc-89}, where $c$ is the central charge and the $\Delta_j$ are the dimensions of the allowed boundary operators consistent with the given boundary conditions on each edge of the annulus. Thus the eigenvalues of the entanglement hamiltonian in all these cases are given, in the scaling limit, by the boundary scaling dimensions of the CFT, as already observed in Refs.~\cite{lauchli,tachikawa-14}. In particular, the methods we use in this paper coincide with those of Ref.~\cite{tachikawa-14} insofar as the examples considered there.

If we compare (\ref{KAgeneral}) with the result for a thermal ensemble
\be\label{Kthermal}
K_{\textrm{\tiny thermal}}
=\frac\beta{2\pi}\int_AT_{00}(x)dx\,,
\ee
 we see that $f'(x)^{-1}$ may be interpreted as a position-dependent effective inverse temperature
\be
\label{beta eff def}
\beta_{\textrm{\tiny eff}}(x)=\frac{2\pi}{f'(x)}\,.
\ee
In the case when $\beta$ is uniform the thermal entropy per unit length is \cite{bcn-86,aff-86}
\be
s_{\textrm{\tiny thermal}} = \frac{\pi c}{3\beta}\,,
\ee
and so one might be tempted \cite{klich-13} to identify the entanglement entropy from (\ref{beta eff def}) as
\be
\label{SA?}
S_A\stackrel{?}{=}\,\frac{\pi c}{3}\int_A\beta_{\textrm{\tiny eff}}(x)^{-1}dx=\frac c6\int_Af'(x)dx=\frac{c}{6}\, W\,.
\ee
Since $f(x)\sim\log(x-x_j)$ at each entangling point $x_j$, this expression diverges as $\sim(c/6)\log(1/\epsilon)$
at each such point, as expected. 
In fact (\ref{SA?}) is correct up to $O(1)$ terms, but for the wrong reason. Entanglement entropy, unlike the Gibbs entropy, is not in general an extensive quantity.

A correct argument is as follows.
As we shall show, the R\'enyi entropies are proportional to a ratio of partition functions on the annulus, and in the limit $W\gg1$ this is dominated by the ground state $|0\rangle$ of the hamiltonian which generates translations in the direction of Re $w$:
\be
{\rm Tr}\,\rho^n_A\sim(\langle a|0\rangle \langle0|b\rangle)^{n-1}\,e^{-\frac{c}{12}(n-1/n)W}\,,
\ee
where $a,b$ denote the boundary states. This implies that
\be
S_A\sim\frac{c}{6}\,W+g_a+g_b\,,
\ee
where $g_{a,b}=-\log\langle a,b|0\rangle$ are the Affleck-Ludwig boundary entropies. The existence of such terms
for external boundaries was deduced in Ref.~\cite{cc-04}, but we see that they also originate from the boundary conditions at the entangling points. 

We see from this that (\ref{SA?}) arises as a consequence of the particular properties of 2d CFTs, and there is no reason to expect it to hold, for example, in higher dimensions, except possibly as the R\'enyi index $n\to0^+$ \cite{swingle-16}.

\begin{figure}
\vspace{.1cm}
\includegraphics[width=1\textwidth]{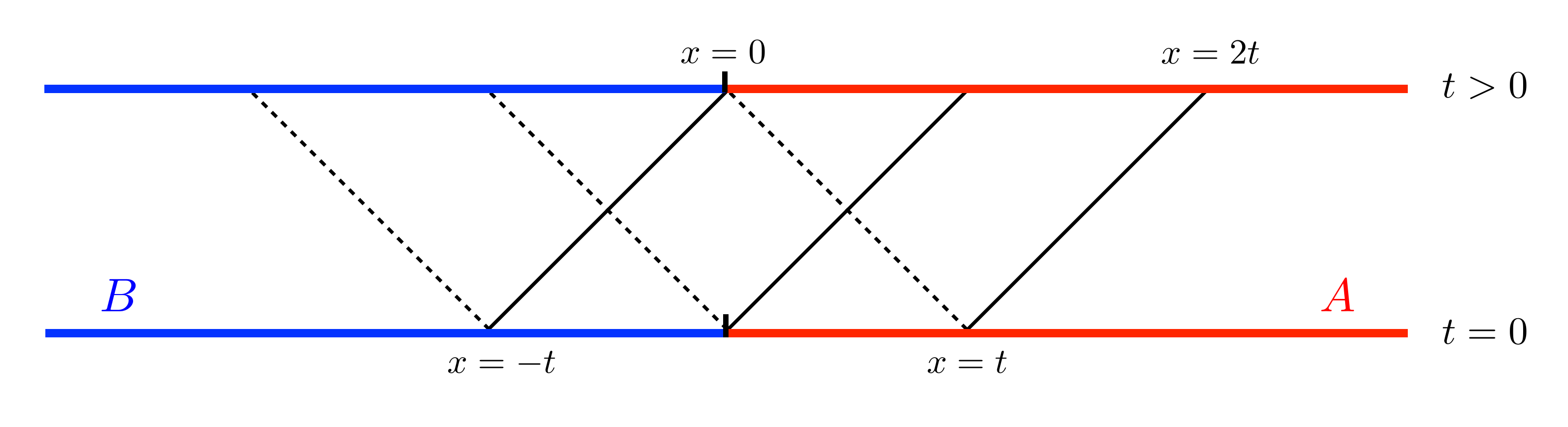}
\vspace{-.7cm}
\caption{Quasiparticle picture of the hamiltonian $K_A$ describing the entanglement between the semi-infinite intervals $A$ and $B$, after a global quench, as given in (\ref{Kglobalapprox}). Most of the entanglement is thermal, due to the R-moving particles (shown as solid lines) of pairs emitted from the interval $(-t,t)$, and reaching the subinterval $(0,2t)$ of $A$ at time $t$. The L-movers (shown as dashed lines) are correlated with these, and contribute similarly to $K_B$. 
} 
\label{fig:qp global}
\end{figure}

The new examples in this paper correspond to time-dependent situations commonly termed quantum quenches. We treat both the cases of a global and local quench. In the global case, we consider the time-evolution in the CFT of 
an initial state $e^{-(\beta/4)H_{\textrm{\tiny CFT}}}|b\rangle$ where $|b\rangle$ is a conformally invariant boundary state \cite{cc-05-global quench}. 
While this appears to be rather special, it has the properties of initial short-range correlations and entanglement, and also that of thermalization, in that after a time $t\sim\ell/2$ the reduced density matrix $\rho_A$ of a subsystem of length $\ell$ becomes indistinguishable from that of a thermal ensemble at temperature $\beta^{-1}$ \cite{jc-15 gge}. 
We consider the entanglement between the semi-infinite regions $A=(0,\infty)$ and $B=(-\infty,0)$.
In that case it was shown in Ref.~\cite{cc-05-global quench} that the R\'enyi entropies are given by the euclidean path integral on the $n$-sheeted cover of an infinitely long strip, with a single twist operator inserted at $\textrm{i} \tau$ such that $|\tau|<\beta/4$, analytically continued to $\tau\to \textrm{i}t$. In this case there are two boundaries, one around the entangling point and the other corresponding to $|b\rangle$. At $t=0$ we then find
\be
\label{Kglobal t=0}
K_A=\frac{\beta}{2\pi}\int_0^\infty\sinh(2\pi x/\beta) \, T_{00}(x)dx\,.
\ee
For $x\ll\beta$ the weight factor agrees with (\ref{halfspace}), but for $x\gg\beta$ it becomes exponentially large, corresponding to very low effective temperature and a very small contribution to the entanglement entropy. On the other hand, for $t\gg\beta$ we find
\be
\label{Kglobalapprox}
K_A\approx \frac\beta\pi\int_{0}^{2t}T(x,t)dx+\cdots=\frac\beta\pi\int_{-t}^tT(x,0)dx+\cdots\,,
\ee
where $T(x,t)$ is the energy-momentum tensor for only the right-movers (and is in fact a function of only $x-t$).
This approximation simplifies the behavior near the light-cones $|x|=t+O(\beta)$, and suppresses a contribution to short-range entanglement for $x>t$.  
(For the complete expression see Eqs.\,(\ref{K_A global time v1}) and (\ref{K_A global time v2}).) 
This result has two new features. First, it depends only on the stress tensor $T$ for the R-movers, which equals
$\frac12(T_{00}+T_{10})$, thus involving not only the hamiltonian density $T_{00}$ but also the momentum density $T_{10}$. Second, although the first expression is an integral along $A$ at the observation time $t$, this is equivalent to an integral along the intersection of the past light-cone of $A$ with any constant time slice, in particular $t=0$. However it is consistent with the quasi-particle picture of the global quench put forward in Ref.~\cite{cc-05-global quench}, in which the entanglement between $A$ and $B$ is due to EPR pairs, entangled over a spatial distance $O(\beta)$, emitted from the interval $(-t,t)$ of the initial state, as illustrated in Fig.~\ref{fig:qp global}. We now see that the entanglement of $A$ with $B$ in fact comes only from the R-movers (while a similar expression for $K_B$ would involve only the left-movers). There is no discrepancy here, since the L-movers are correlated with the R-movers over a distance $O(\beta)$ due to the conformal boundary condition $T=\overline T$ at $\tau=-\beta/4$. The entanglement entropy computed from (\ref{Kglobalapprox}) is $S_A(t)\sim(\pi c/3\beta)t$, in agreement with Ref.~\cite{cc-05-global quench}, and the spectrum of $K_A$ is again that of a boundary CFT,
with spacing $\sim\beta\Delta_j/2t$.

\begin{figure}
\vspace{.1cm}
\includegraphics[width=1\textwidth]{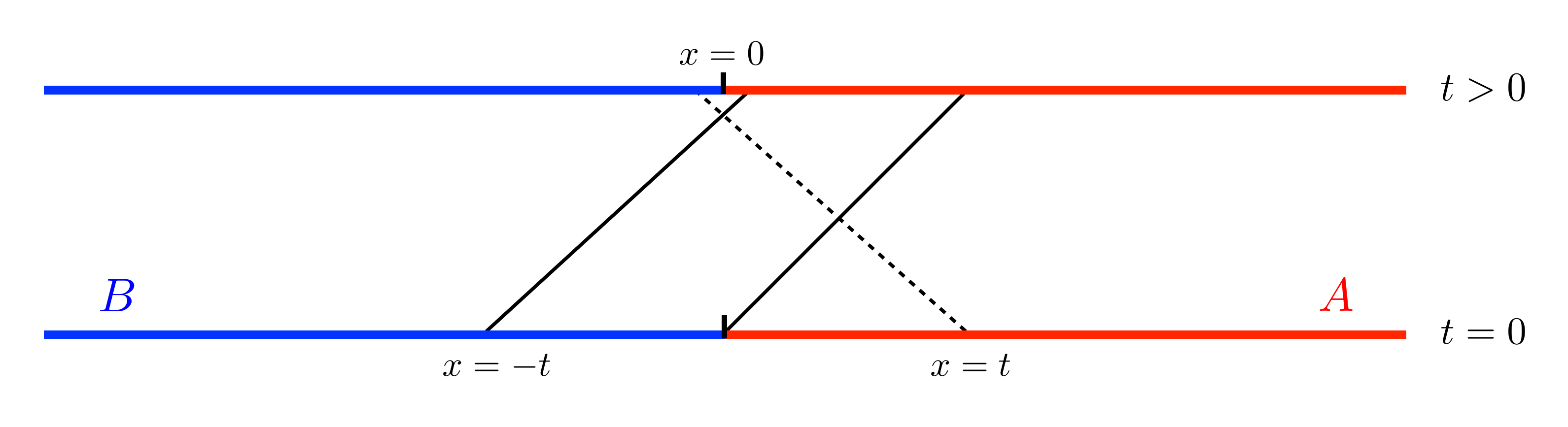}
\vspace{-.7cm}
\caption{Quasiparticle picture of the hamiltonian $K_A$ describing the entanglement between the semi-infinite intervals $A$ and $B$, after a local quench where two semi-infinite systems are joined together at $x=0$ at time $t=0$, as given in (\ref{KA local large t}). Most of the entanglement comes from R-moving particles of pairs emitted from near the junction, but also from R- and L-movers from near $x=-t$ and $x=t$ respectively.} 
\label{fig:qp local}
\end{figure}

For the local quench, we consider the case when two semi-infinite intervals, each in their ground state, are joined together smoothly at $t=0$, and the subsequent evolution is by the $H_{\textrm{\tiny CFT}}$ on the full line. We  treat various cases, but again the simplest is to study the entanglement between $A=(0,\infty)$ and $B=(-\infty,0)$. We find, again for late times,
\be
\label{KA local large t}
K_A\approx\int_{-t}^\infty\frac{(x+t)|x|}{t}\, T(x,0)dx +\int_t^\infty\frac{(x-t)|x|}{t}\, \overline T(x,0)dx\,.
\ee
(This expression has again been simplified around $x\approx \pm t$ and $x\approx 0$. See Eqs.\,(\ref{K_A local t}) and (\ref{K_A local t v2}) for the full result.) 
Once again we see that the L- and R-movers enter differently, and that the integral at $t=0$ is over the past light-cone of $A$. The main contributions to the entanglement entropy however, come from the regions $x\approx\pm t$ and $x\approx 0$, as illustrated in Fig.~\ref{fig:qp local}. 
These contribute differently to the entropy (see (\ref{cal W local}) and (\ref{cal bW local})), but in total
we find $S_A(t)\sim(c/3)\log t$, in agreement with Ref.~\cite{cc-07-local quench}. The spectrum of $K_A$ is again that of a boundary CFT, now with spacing $\sim\pi\Delta_j/(2\log t)$.
\vspace{5mm}

The outline of this paper is as follows: in Sec.~\ref{sec2} we discuss the mapping to the annulus in general and show how this determines the entanglement spectrum and the entanglement entropy. In Sec.~\ref{sec3} we apply this to the static examples previously discussed in the literature, 
adding a new one.
Then, in Sec.~\ref{sec4} we treat the cases of a global and local quench in detail. We also summarize results for an inhomogenous quench. Finally in Sec.~\ref{sec5} we summarize and discuss possible partial extensions of these results to higher genus and higher dimensions. 

\section{Mapping to the annulus}
\label{sec2}

As usual, the reduced density matrix $\rho_A$ may be computed as the path integral on a suitable euclidean space-time, cut open along the intersection $C$ of the spatial region $A$ with a constant time slice. For the static cases we consider (either the vacuum state or finite temperature), it does not matter at which time slice $C$ is placed, and we may take it to be at euclidean time $\tau=0$. In the time-dependent examples in Sec.~\ref{sec4}, we choose it to lie at euclidean time $\tau$, and then continue $\tau\to\textrm{i}t$. 

The R\'enyi entropies are then given by sewing together $n$ copies of this path integral cyclically along $C$, giving an $n$-sheeted cover of the original euclidean space-time, with conical singularities at the entangling points, which are the ends of $C$. Denoting the partition function on this $n$-sheeted cover by $Z_n$, we then have
\be\label{ratio}
{\rm Tr}_{{\cal H}_A}\rho_A^n=\frac{Z_n}{Z_1^n}\,.
\ee

In all the examples we consider both the original space and its $n$-sheeted cover are conformally equivalent to an annulus, once proper account has been taken of the regularization by removing a region around the entangling points. In the introduction, we argued that this may be done by removing small slit along $\tau=0$ in the euclidean path integral, but it is more convenient to remove a small disc since this regulator respects the local rotational invariance. This does not affect the topology, and should not affect any universal properties of the result (such as the contributions of the boundary entropy), but it will affect the non-universal amplitudes of corrections to scaling. 

Having done this, let $z\to w=u+\textrm{i}v=f(z)$ be the conformal mapping of the single cover to the annulus. This is normalized in such a way that the boundary around one of the entangling points is mapped into $u=$ constant, 
$0\leqslant v<2\pi$, periodically identified mod$(2\pi)$. Thus $f(z)\sim\log(z-z_j)$ near this point. The $n$-sheeted cover is then mapped into an annulus with $v$ periodically identified mod$(2\pi n)$. In the static cases, it will turn out that $C$ is mapped into a constant $v$ section of the annulus, which we may take to be $v=0$, but in the time-dependent cases (before continuing to real time) it is mapped to a more general curve connecting the two boundaries of the annulus. 

The entanglement hamiltonian $K_A$ is then the conformal image of the generator of translations around the annulus in the direction of $v={\rm Im}\,w$:\footnote{The hamiltonian density is $T_{00}$ in Minkowski signature, which is $-T_{vv}$ in euclidean space. In complex coordinates this is $T+\overline T$.}
\be
K_A=-\int_{v={\rm const.}}T_{vv}du=\int_{f(C)}T(w)dw+\int_{\overline{f(C)}}\overline T(\bar w)d\bar w\,,
\ee
which becomes
\be\label{KAgeneral T barT}
K_A=\int_C\frac{T(z)}{f'(z)}\,dz+ \int_{\overline C}\frac{\overline T(\bar z)}{\overline{f'(z)}}\,d\bar z\,.
\ee
The factor of $f'(z)^{-1}$ in the first term is a product of the $f'(z)^{-2}$ occurring in the transformation rule for $T$, and a 
jacobian factor $f'(z)$. We have ignored the Schwartzian term in the transformation law since this gives a constant which cancels in the ratio (\ref{ratio}). (However, it needs to be included if we insist that ${\rm Tr}\,e^{-2\pi K_A}=1$.)

The entanglement hamiltonian $K_A$ is locally the generator of rotations around the entangling point(s). Its exponentiation $e^{-2\pi K_A}$ covers the euclidean space-time, minus the regularizing discs, exactly once:
\be\label{ZnZ1}
Z_1={\rm Tr}\,e^{-2\pi K_A}\,,\qquad Z_n={\rm Tr}\,e^{-2\pi n K_A}\,.
\ee

However, we may equally well compute these partition functions on the annulus $\{w\}$. Denoting the width of the annulus by $W$, the eigenvalues of $K_A$ are, apart from a constant, then given by $\pi(-c/24+\Delta_j)/W$, where the $\Delta_j$ are dimensions of the allowed boundary operators consistent with the prescribed boundary conditions on each boundary \cite{jc-89}.
 
 It is useful to define the modular parameters
\be
q\equiv e^{-2\pi^2/W}\,,\qquad \tilde q=e^{-2W}\,,
\ee
in terms of which \cite{jc-89}
\be\label{Zq}
Z_1=q^{-c/24}\sum_jd_j \, q^{\Delta_j}\,,\quad Z_n=q^{-nc/24}\sum_jd_j \,q^{n\Delta_j}\,,
\ee
where the positive integers $d_j$ are degeneracy factors.

Equivalently, by considering the generator of translations in the $u$-direction \cite{jc-89}
\be\label{Ztildeq}
Z_1=\tilde q^{-c/24}\sum_k\langle a|k\rangle\langle k|b\rangle \, \tilde q^{\delta_k}\,,
\quad Z_n=\tilde q^{-c/24n}\sum_k\langle a|k\rangle\langle k|b\rangle \,  \tilde q^{\delta_k/n}\,,
\ee
where the sum is now over all allowed scalar bulk operators with dimensions $\delta_k$, and $|a,b\rangle$ denote the boundary states. In fact, since
$W={\rm Re} \int_{f(C)}dw={\rm Re}\int_Cf'(z)dz$, $W\propto\log(1/\epsilon)\gg1$, so that $q\approx 1$,
$\tilde q\ll1$ in the region of interest, and the expressions in (\ref{Ztildeq}) are more useful than those in (\ref{Zq}).

From (\ref{ZnZ1}), (\ref{Zq}) and (\ref{Ztildeq}) we then see that the eigenvalues of $\rho_A$, properly normalized so that ${\rm Tr}\,\rho_A=1$, are of the form
\be 
\frac{q^{-c/24+\Delta_j}}{Z_1(q)}\sim\frac{q^{-c/24+\Delta_j}}{\langle a|0\rangle\langle 0|b\rangle \, \tilde q^{-c/24}}\,,
\ee
with degeneracies $d_j$. From this may be read off the universal part of the spectrum of $K_A=-(1/2\pi)\log\rho_A$.
We stress that this is completely universal across all the examples considered in the subsequent sections, the only difference being in how $W$, and therefore $q$ and $\tilde q$, depend on the geometry, and the chosen boundary conditions. 
In particular, we see that the smallest eigenvalue of $-\log\rho_A$, corresponding to $\Delta_j=0$, is
\be\label{lambdamax}
\lambda_{\textrm{\tiny min}}\sim\log\left(\langle a|0\rangle\langle 0|b\rangle \, \tilde q^{-c/24}\right)=
\frac c{12}W-g_a-g_b\,,
\ee
where $g_{a,b}=-\log\langle a,b|0\rangle$ are the boundary entropies.

We also see from (\ref{ratio}) and (\ref{Ztildeq}) that, for $W\gg1$,
\be\label{Wbig}
{\rm Tr}\,\rho_A^n\sim \frac{\langle a|0\rangle\langle 0|b\rangle \, \tilde q^{-c/24n}}
{\big(\langle a|0\rangle\langle 0|b\rangle\big)^n \, \tilde q^{-cn/24}}\,,
\ee
so that the R\'enyi entropies are
\be
\label{SA_n generic}
S_A^{(n)}\sim \frac c{12}\left(1+\frac1n\right)W+g_a+g_b\,.
\ee
The `unusual' corrections to this are  powers of $\tilde q^{\delta_j/n}$ and $\tilde q^{\delta_j}$, as predicted in Ref.~\cite{cc-10}.
Note the leading term in the $n=1$ entanglement entropy is twice that in the smallest eigenvalue of $-\log\rho_A$ in (\ref{lambdamax}), as observed in \cite{calabrese-lefevre}, but this is no longer the case once the $O(1)$ boundary entropy terms are included.
In this reference the density of states of $K_A$ was found by transforming the leading term given by $(\ref{Wbig})$: this gives the approximate form, but we see that in fact the true density of states is just that of a (boundary) CFT.

For minimal models, or more generally rational CFTs, one may say more \cite{jc-89}. The sums in (\ref{Zq}) and (\ref{Ztildeq})
may be organized into finite sums over characters $\chi_j(q)=q^{-c/24+\Delta_j}\sum_N d_{j,N} \, q^N$:
\be
Z_n=\sum_j  n_j \, \chi_j(q^n)=\sum_{j,k}n_j \, S_j^k \,\chi_k(\tilde q^{1/n})\,,
\ee
where the non-negative integers $n_j$ provide the operator content given by the boundary conditions chosen on the annulus, and $S_j^k$ are the elements of the modular $S$-matrix. Thus
\be
{\rm Tr}\,\rho^n_A
=\frac{\sum_j n_j \, \chi_j(q^n)}{\big(\sum_jn_j \, \chi_j(q)\big)^n}
=\frac{\sum_{j,k}n_j \, S^k_j \, \chi_k(\tilde q^{1/n})}{\big(\sum_{j,k} n_j \, S^k_j \, \chi_k(\tilde q)\big)^n}\,.
\ee
Interestingly, this has the same form as is found for the reduced density matrix of a semi-infinite interval in a non-critical integrable lattice model using the corner transfer matrix, with, however, a different physical meaning for 
$q$ and $\tilde q$ \cite{ccp-10}. In particular in this case $\tilde q\sim\xi^{-2}$ as the correlation length $\xi\to\infty$.
(See \cite{katsura} for recent numerical results.)

\section{Time-independent examples}
\label{sec3}

\subsection{Single interval on the infinite line, ground state}
\label{sec3a}

\begin{figure}
\vspace{.3cm}
\includegraphics[width=1.\textwidth]{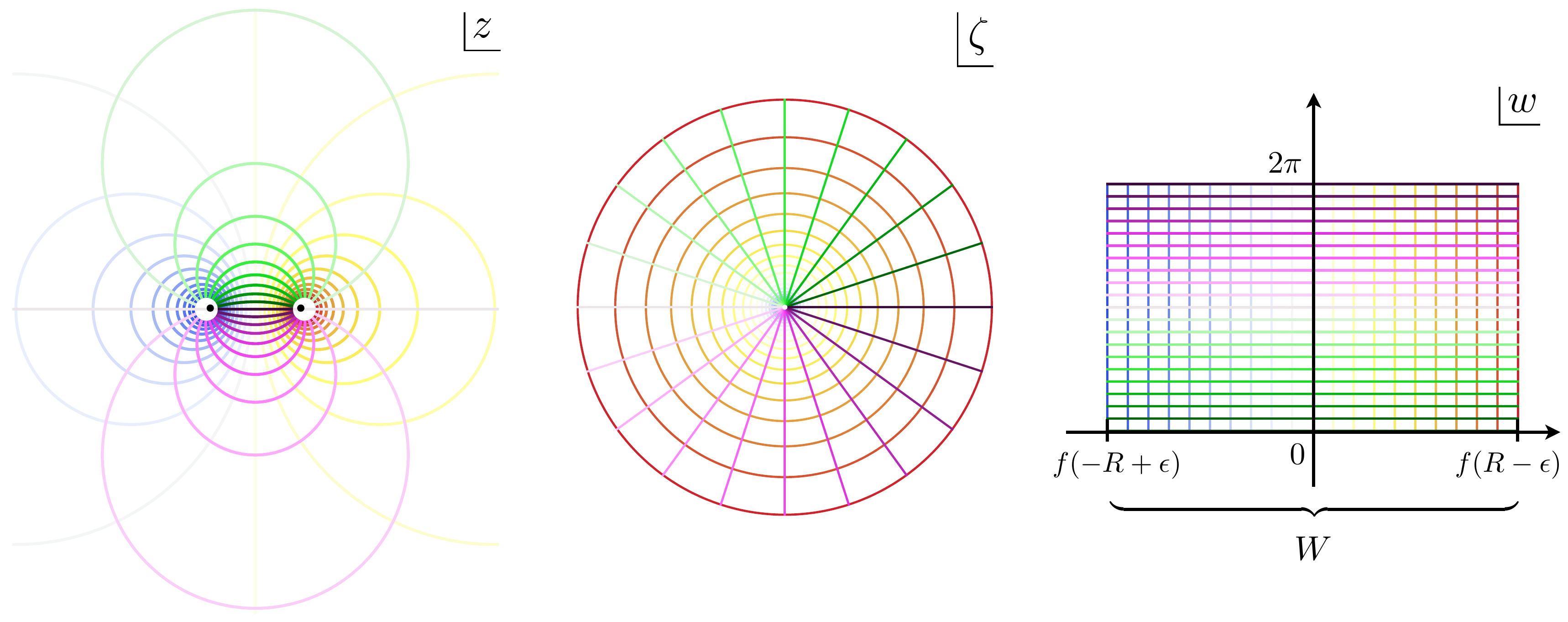}
\vspace{-.4cm}
\caption{
Euclidean space-times characterising the case of a single interval $A=(-R,R)$ on the infinite line, as discussed in Sec.\,\ref{sec3a}. 
In the left panel, the black dots denote the endpoints of the interval $A$.
In the right panel the segments with $\textrm{Im}(w) =0$ and $\textrm{Im}(w) =2\pi$ are identified.
} 
\label{fig:interval infinite line T=0}
\end{figure}

Suppose $A$ is the interval $(-R,+R)$, and the whole system is in the ground state. The euclidean space-time is then
the complex $z$-plane, minus small discs of radius $\epsilon$ around the endpoints at $\pm R$.

The mapping to the annulus parameterized by the complex coordinate $w=f(z)$ can be constructed by first mapping the $z$-plane into the domain   delimited by the images of the two small discs above (described by the complex coordinate $\zeta$) and then mapping it onto the annulus given by $w=\log(\zeta)$, being $\zeta = \tfrac{z+R}{R-z}$. 
The complete map reads
\be
\label{fmap single interval}
w=f(z)=\log\left(\frac{z+R}{R-z}\right) .
\ee
In Fig.\,\ref{fig:interval infinite line T=0} we show the euclidean space-times involved in the construction of the map (\ref{fmap single interval}).

Note that $f(z)$ is just the (complex) electrostatic potential due to charges $\pm1$ at the entangling points. The curves 
${\rm Re}\,f=$ constant, which map onto $u=$ constant on the annulus, are the equipotentials, while the curves ${\rm Im}\,f=$ constant, which map onto $v=$ constant, are the field lines. This electrostatic analogy is common to all the examples we consider which may be mapped to an annulus.

In the coordinates $w$, the width of the annulus is $W = f(R-\epsilon) - f(-R+\epsilon)= 2\log(\ell/\epsilon) + O(\epsilon)$, where $\ell = 2R$ is the length of $A$.
Thus, from (\ref{SA_n generic}), we find that in this case the R\'enyi entropies have the form
\be
S_A^{(n)} = \frac{c}{6} \left(1+\frac1n\right)\log(\ell/\epsilon)+g_a+g_b+{\rm corrections}\,.
\ee
The leading term is well-known, but the $O(1)$ terms less so. Note that they cannot, for all $n$, simply be absorbed into a redefinition of the regulator $\epsilon$. The boundary entropies are therefore in principle measurable by comparing different values of $n$. If we use the same prescription to regularize around the two entangling points, then $g_a=g_b$.

The entanglement hamiltonian follows from Eq.\,(\ref{KAgeneral}), giving
the well-known form (\ref{KAball}), specialized to 1+1 dimensions.

\subsection{Finite interval in an infinite system at finite temperature}
\label{subsec:finite T}

The finite temperature case was considered in Refs.\,\cite{borchers-98, klich-13}. 
If $A$ is the interval $(-R,+R)$ of length $\ell = 2R$ in the infinite line, this time on a cylinder of circumference $\beta$ in the imaginary time ${\rm Im}(z)$ direction, the conformal mapping $z \to e^{2\pi z/\beta}$ sends this problem into the one considered in Sec.\,\ref{sec3a}. 
The entangling points $z=-R$ and $z=R$ are mapped into $e^{-2\pi R/\beta}$ and $e^{2\pi R/\beta}$ respectively, and so the mapping to the annulus
is
\be
f(z)=\log\left(\frac{e^{2\pi z/\beta}-e^{-2\pi R/\beta}}{e^{2\pi R/\beta}-e^{2\pi z/\beta}}\right) .
\ee
From (\ref{KAgeneral}), the entanglement hamiltonian is now
\be
K_A=
\frac{\beta}{\pi} \int_A
\frac{\sinh[\pi(R-x)/\beta] \, \sinh[\pi(x+R)/\beta]}{\sinh(2\pi R/\beta)}\; T_{00}(x)\, dx\,.
\ee
The width $W$ of the annulus in this case reads
\be
W
=  f(R-\epsilon)-f(-R+\epsilon)
=
 2  \log\bigg(
\frac{\beta}{\pi \epsilon} \sinh(\pi \ell/\beta)
\bigg)
+ O(\epsilon)\,.
\ee
Inserting this result into (\ref{SA_n generic}) we find the R\'enyi entropy
\be
S_A^{(n)}=\frac{c}{6}\left(1+\frac1n\right)
\log\bigg(
\frac{\beta}{\pi \epsilon} \sinh(\pi \ell/\beta)
\bigg)
+g_a+g_b+{\rm corrections}\,,
\ee
in agreement with Ref.\,\cite{cc-04}, up to the boundary terms.

\subsection{Finite interval in a finite system, ground state}

Let us consider a finite interval $A=(-R,+R)$ of length $\ell =2R$ in a finite spatial circle of circumference $L$.
This setup was also addressed in Ref.\,\cite{klich-13}.
Now the euclidean space-time described by $z$ is the cylinder given by $ 0\leqslant \textrm{Re}(z)< L$ and $\textrm{Im}(z) \in \mathbb{R}$, where the two lines $\textrm{Re}(z)=0$ and $\textrm{Re}(z)=L$ are identified.
This problem can be sent into the one treated in Sec.\,\ref{sec3a} by employing the conformal map $z \to e^{2\pi \textrm{i} z/L}$.
Thus, the mapping to the annulus becomes
\be
f(z) =
\log\left( \frac{e^{2\pi \textrm{i} z/L} - e^{-2\pi \textrm{i}  R/L}}{e^{2\pi \textrm{i}  R/L} - e^{2\pi \textrm{i}  z/L}} \right) .
\ee
By specifying (\ref{KAgeneral}) to this case, one gets the entanglement hamiltonian 
\be
K_A=
\frac{L}{\pi} \int_A
\frac{\sin[\pi(R-x)/L] \, \sin[\pi(x+R)/L]}{\sin(2\pi R/L)}\; T_{00}(x)\, dx \,.
\ee
As for the R\'enyi entropies, we need to compute the width of the annulus
\be
W = f(R-\epsilon)-f(-R+\epsilon)
=
 2  \log\bigg(
\frac{L}{\pi \epsilon} \sin(\pi \ell/L)
\bigg)
+ O(\epsilon)\,.
\ee
Then, Eq.\,(\ref{SA_n generic}) tells us that the R\'enyi entropies are
\be
S_A^{(n)} =\frac{c}{6}\left(1+\frac1n\right)
\log\bigg(
\frac{L}{\pi \epsilon} \sin(\pi \ell/L)
\bigg)
+g_a+g_b+{\rm corrections}\,,
\ee
which is the result obtained in Ref.\,\cite{cc-04}, up to the boundary terms.

\subsection{Interval at the end of a semi-infinite line, ground state}
\label{subsec:bdy}

\begin{figure}
\vspace{.3cm}
\includegraphics[width=1.\textwidth]{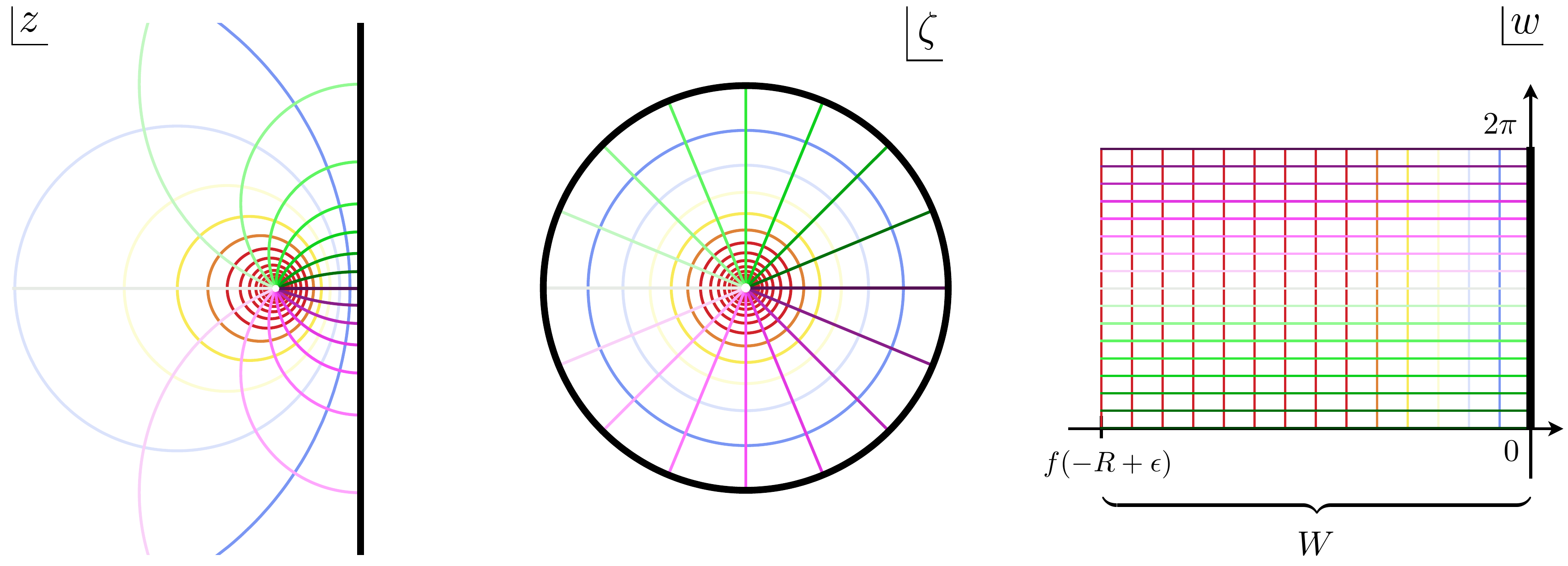}
\vspace{-.3cm}
\caption{
Euclidean space-times describing the case of an interval $A=(-R,0)$ at the end of semi-infinite line, as discussed in Sec.\,\ref{subsec:bdy}.
The thick black lines in the three panels are mapped into each other.
In the right panel the segments on the lines $\textrm{Im}(w) =0$ and $\textrm{Im}(w) =2\pi$ are identified.
} 
\label{fig:interval semi-infinite line T=0}
\end{figure}

In this example $A=(-R,0)$ and $B=(-\infty, -R)$,  being $R>0$ the length of the interval. 
The mapping $f(z)$ to the annulus is again given by (\ref{fmap single interval}), but now one of the boundaries ($b$) is the conformal image of the external boundary along ${\rm Re}\,z=0$.
The euclidean space-times characterising this case are represented in Fig.\,\ref{fig:interval semi-infinite line T=0}, where the black lines, which correspond to the external boundary, are mapped into each other.

In this case the width of the annulus is simply $W= \log(2R/\epsilon) +O(\epsilon)$, and so, from (\ref{SA_n generic}) we get
\be
S_A^{(n)} = \frac c{12}\left(1+\frac1n\right)\log(2R/\epsilon)+g_a+g_b+{\rm corrections}\,,
\ee
where now we may have $g_a\not=g_b$. 

The entanglement hamiltonian has the same form given by (\ref{KAball}) specialized to 1+1 dimensions, integrated over the interval $A$.

\subsection{Finite interval at the end of a segment with the same b.c., ground state}
\label{sec:2bdy}

Consider now the case where the whole system $A\cup B$ is given by the interval $(-L/2,+L/2)$. 
The conformal boundary conditions at ${\rm Re}\,z=\pm L/2$ are taken to be the same (see below for the more general case.) The subsystem $A$ is the interval $(x_0,L/2)$. 
As usual, we remove a small disc around the entangling point $z=L/2-\ell$. The remaining region is once again conformally an annulus.

The conformal map which takes this to the conventional presentation of the annulus in the $w$-plane is
\be
\label{logsincos}
w=f(z)=\log\left(\frac{\sin[\pi(z-x_0)/2L]}{\cos[\pi(z+x_0)/2L]}\right) .
\ee
This may be constructed, using the method of images, as the complex electrostatic potential due to a unit charge at $z=x_0$ between two conducting plates at ${\rm Re}\,z=\pm L/2$. In any case, it may easily be checked that, when ${\rm Re}\,z=\pm L/2$, ${\rm Re}\,f(z)=0$, and that when $z$ is real, so is $f(z)$. 
An alternative derivation is also given in Appendix~\ref{app: maps}.

We then have 
\be
f'(z)=
\frac{\pi}{2L} \, \big(\cot [\pi(z-x_0)/2L]+\tan[\pi(z+x_0)/2L] \,\big)
=\frac\pi L\left(\frac{\cos(\pi x_0/L)}{\sin(\pi z/L)-\sin(\pi x_0/L)}\right),
\ee
so that Eq.\,(\ref{KAgeneral}) adapted to this case provides the entanglement hamiltonian 
\be
\label{K_A segment same bc}
K_A=
\frac{L}{\pi} \int_A
\frac{\sin(\pi x/L) - \sin(\pi x_0/L)}{\cos(\pi x_0 /L)}\; T_{00}(x)\, dx\,,
\ee
where the integration domain is $A=(x_0 +\epsilon, L/2)$.
Notice that, considering the regime $\ell/L \ll 1$ in (\ref{K_A segment same bc}), the expected result derived in Sec.\,\ref{subsec:bdy} is recovered. 

In this case  the width $W$ of the annulus is
\be
W = f(L/2) - f(x_0+\epsilon) = 
  \log\bigg(
\frac{2L}{\pi \epsilon} \sin(\pi \ell/L)
\bigg)
+ O(\epsilon)\,, 
\ee
where $\ell=L/2-x_0$.
Thus, from (\ref{SA_n generic}), the R\'enyi entropies $S_A^{(n)}$ read
\be\label{SA_n seg same bc}
S_A^{(n)} = \frac{c}{12}\left(1+\frac1n\right)   
\log\bigg( \frac{2L}{\pi \epsilon} \sin(\pi \ell/L) \bigg)
+g_a+g_b + \textrm{corrections}\,,
\ee
where $g_a$ comes from the conformal boundary state around the branch point, while $g_b$ encodes the conformal boundary state on the lines $\textrm{Re}(z)=\pm L/2$. 

The case where the boundary conditions are different on each boundary at $x=\pm L/2$ is much more difficult. This is because the map (\ref{logsincos}) now introduces $2n$ boundary-condition changing operators on one of the boundaries of the annulus, at the points $w=\textrm{i}\pi(\frac12+ \frac{x_0}L)+2\textrm{i}\pi k$ and
$\textrm{i}\pi(\frac32- \frac{x_0}L)+2\textrm{i}\pi k$, where $k=0,\ldots,n-1$. In this case, since the width $W$ of the annulus is the same, the leading term in (\ref{SA_n seg same bc}) is unchanged, as is the term $g_a$, but the $O(1)$ term $g_b$ is replaced by a universal but complicated function of $n$ and $x_0/L$. In the limits $x_0\to\pm L/2$, however, it should approach the entropy of the appropriate boundary.

\section{Quantum quenches}
\label{sec4}

In this section we consider the temporal evolution of the entanglement hamiltonians of two dimensional CFTs after some quantum quenches.
A quantum quench is a sudden change in one of the quantities determining the unitary temporal evolution of the initial state. 
In particular, we focus on global quenches \cite{cc-05-global quench}, local quenches \cite{cc-07-local quench}, and
inhomogeneous quenches \cite{sot-jc-08}. For a more complete list of references 
we refer the reader to the recent review \cite{cc-rev-quench}.

\subsection{Global quench}
\label{sec:global quench}

In a (homogeneous) global quench, a system is prepared at time $t=0$ in a translationally invariant state $|\psi_0\rangle$, which may be the ground state of some hamiltonian $H_0$, and is subsequently unitarily evolved with a different hamiltonian $H$, of which
 $|\psi_0\rangle$ is not an eigenstate, and with respect to which it has an extensively large energy. 

This problem, when $H$ is the hamiltonian of a CFT, was first addressed in \cite{cc-05-global quench}, where the time evolution of the entanglement entropy $S_A(t)$ of an interval was described. Subsequently the time evolution of the correlation functions \cite{cc-06-cf} and of the reduced density matrix $\rho_A(t)$ \cite{jc-15 gge} were also discussed. In these works a particular form was taken for $|\psi_0\rangle$, namely $\propto e^{-(\beta/4)H}|b\rangle$, where $|b\rangle$ is a conformal boundary state and $\beta$ is a parameter with the dimensions of inverse temperature. 

Such states have short-range correlations and entanglement, and are supposed to be reasonable approximations to the ground states of gapped QFT hamiltonians $H_0$. They possess the technical advantage that the time-evolution $e^{-\textrm{i}Ht}|\psi_0\rangle$ is analytically tractable, and the physical property of subsystem thermalization, that is the reduced density matrix $\rho_A(t)$ of a subsystem of length $\ell$ becomes exponentially close to that of a Gibbs ensemble $e^{-\beta H}$ for times $t>\ell/2$. (More general initial states lead to a generalized Gibbs ensemble (GGE) \cite{jc-15 gge}.) 

\begin{figure}
\vspace{.1cm}
\hspace{-.2cm}
\includegraphics[width=1.\textwidth]{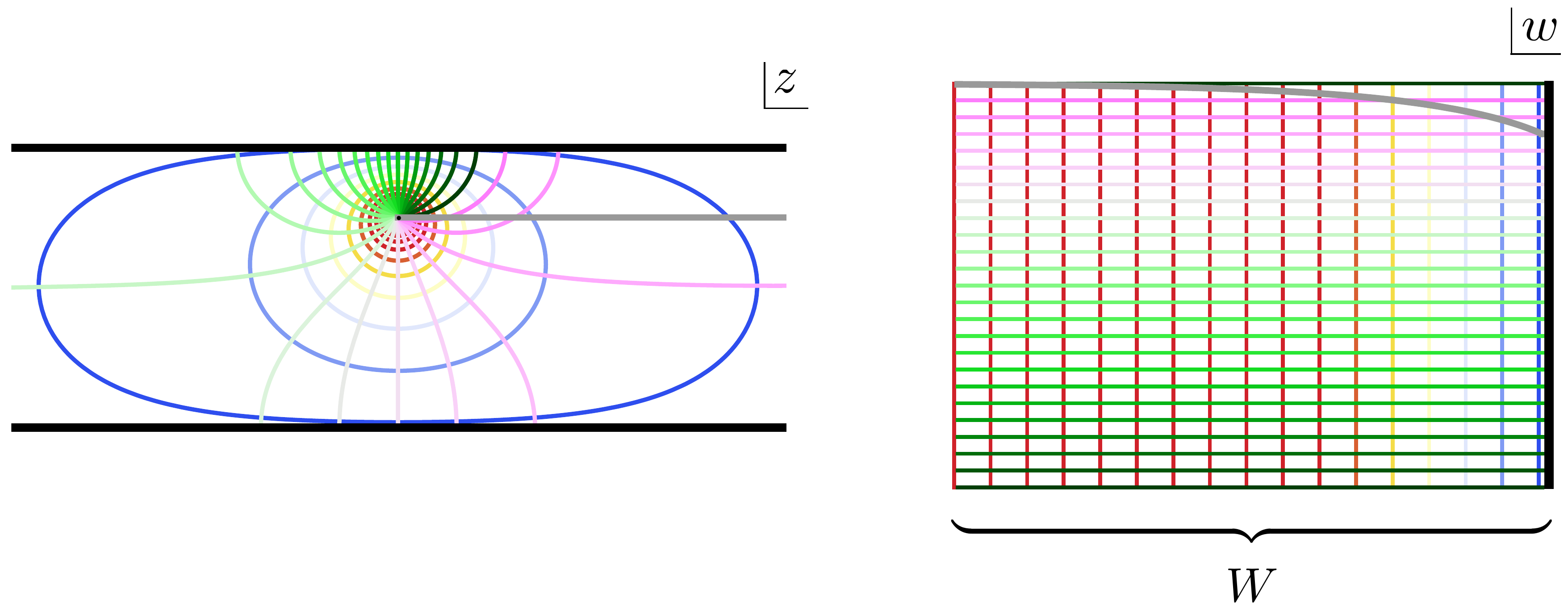}
\vspace{.1cm}
\caption{
Euclidean space-times characterising the case of the semi-infinite line after a global quench, as discussed in Sec.\,\ref{sec:global quench}. Left: The infinite strip whose vertical width is $\beta/2$, which is mapped onto the annulus (right panel) by the conformal map (\ref{fglobal}). The grey line corresponds to the path $C = \{\textrm{i}\tau + x, \, x\geqslant 0\}$. 
Right: The annulus parameterized by the complex coordinate $w$ (the horizontal segments delimiting the rectangle are identified). 
The endpoints of the grey line provide $\mathcal{W}$ in Eq.\,(\ref{calW global}), whose real part gives the width $W$ of the annulus, which leads to the R\'enyi entropies (\ref{SAn time-dep global}). 
} 
\label{fig:global}
\end{figure} 

Let us consider the reduced density matrix $\rho_A$ of a semi-infinite line $x\geqslant 0$.
In the Euclidean setup introduced in \cite{cc-05-global quench}, the global quench can be described by the infinite strip given by $-\beta/4 \leqslant \textrm{Im}(z) \leqslant \beta/4$ and $\textrm{Re}(z)  \in \mathbb{R}$.
Given our choice of $A$, in this space we have to consider $C=\{z=\textrm{i} \tau + x, x\geqslant 0\}$, where $|\tau| < \beta/4$ (see the left panel of Fig.\,\ref{fig:global}, where $C$ is the grey line).
We remark that $C$ and $\overline{C}$ do not coincide in this case. 

As usual, we remove a small disc around the entangling point $z_0=\textrm{i}\tau$, with conformal boundary condition $a$. Since the boundary conditions on ${\rm Im}\,z=\pm\beta/4$ are the same ($b$), the remaining region is once again topologically an annulus. In fact the geometry is a rotated version of that considered in Sec.~\ref{sec:2bdy}, and the conformal mapping to the standard presentation of the annulus is a modification of that in (\ref{logsincos}), namely

\be
\label{fglobal}
w=f(z) \,=\, 
\log\left(\frac{\sinh[\pi(z-\textrm{i}\tau)/\beta]}{\cosh[\pi (z+\textrm{i}\tau)/\beta]} \right) ,
\ee
whose derivative reads
\be
f'(z) = \frac{(2\pi/\beta) \cosh(2\pi \textrm{i}\tau/\beta)}{\sinh(2\pi z/\beta) - \sinh(2\pi \textrm{i}\tau/\beta)}\,.
\ee

By employing this expression in (\ref{KAgeneral T barT}), 
we find for  the entanglement hamiltonian 
\bea
\label{KA global tau}
K_A &=&
\frac{\beta}{2\pi} \int_0^{\infty} 
\frac{\sinh(\pi x/\beta) \, \cosh(\pi [x+2\textrm{i}\tau]/\beta)}{\cosh(2\pi \textrm{i}\tau/\beta)}\;
T(x+\textrm{i}\tau)\, dx
\\
\rule{0pt}{.8cm}
& & +\,
\frac{\beta}{2\pi} \int_0^{\infty} 
\frac{\sinh(\pi x/\beta) \, \cosh(\pi [x-2\textrm{i}\tau]/\beta)}{\cosh(2\pi \textrm{i}\tau/\beta)}\;
\overline{T}(x-\textrm{i}\tau)\, dx\,.
\nonumber
\eea
Notice that the integration domain of these two integrals is the same but the weight factors multiplying $T$ and $\overline{T}$ within the integrals are different. This has its origin in the fact that $C\not=\overline C$.

Making the analytic continuation $\tau\to \textrm{i}t$ of (\ref{KA global tau}), we obtain the time dependent entanglement hamiltonian after the global quench
\bea
\label{K_A global time v1}
K_A &=&
\frac{\beta}{\pi} \int_0^{\infty} \frac{\sinh(\pi x/\beta) \, \cosh(\pi [x-2t]/\beta)}{\cosh(2\pi t/\beta)}\; T(x-t)\, dx
\\
\rule{0pt}{.8cm}
& & +\,
\frac{\beta}{\pi} \int_0^{\infty} \frac{\sinh(\pi x/\beta) \, \cosh(\pi [x+2t]/\beta)}{\cosh(2\pi t/\beta)}\;   \overline{T}(x+t)\, dx \,.
\nonumber
\eea
In this expression it is important to realize that the components of the stress tensor are in fact evaluated at time $t$, that is
\be
T(x,t)=T(x-t)=T(x-t,0)\,,\qquad\overline T(x,t)= \overline T(x+t)=\overline T(x+t,0)\, ,
\ee
as implied by the equations of motion. Thus,
employing $x-t$ as integration variable in the first integral and $x+t$ in the second integral, the entanglement hamiltonian can also be written as
\bea
\label{K_A global time v2}
K_A &=&
\frac{\beta}{\pi} \int_{-t}^{\infty} \frac{\sinh(\pi [x+t]/\beta) \, \cosh(\pi [x-t]/\beta)}{\cosh(2\pi t/\beta)}\; T(x)\, dx
\\
\rule{0pt}{.8cm}
& & +\,
\frac{\beta}{\pi} \int_{t}^{\infty} \frac{\sinh(\pi [x-t]/\beta) \, \cosh(\pi [x+t]/\beta)}{\cosh(2\pi t/\beta)}\; \overline{T}(x)\, dx \,.
\nonumber
\eea
where we should now think of $(T,\overline T)$ as evaluated at $t=0$. 

The formula (\ref{K_A global time v1}), or equivalently (\ref{K_A global time v2}), displays the features anticipated in Sec.\,\ref{sec:intro} for the global quench case.
Let us consider some interesting regimes for this result. 

For $t=0$, the entanglement hamiltonian simplifies to (\ref{Kglobal t=0}), given that $T+  \overline{T} = T_{00}$.

When $t \gg \beta$, from (\ref{K_A global time v2}) we find (\ref{Kglobalapprox}), and that the antiholomorphic part does not contribute in this regime. This approximation ignores interesting contributions from $x+t=O(\beta)$ in the first term, and $x-t=O(\beta)$ in the second. Together, they appear in (\ref{K_A global time v1}) in the form $\int_\epsilon^{O(\beta)}x\,T_{00}(x) dx$, as must happen close to the entangling point.

Another interesting limit is $\beta \to \infty$, which projects $|\psi_0\rangle$ onto the ground state. Then (\ref{K_A global time v1}) becomes
\be\label{betainfty}
\lim_{\beta \to \infty} K_A  =  \int_{0}^{\infty} x\big(T(x-t)+\overline{T}(x+t)\big)\, dx=\int_{0}^{\infty} x\,T_{00}(x,t)\,dx\,.
\ee
Thus (\ref{halfspace}) is recovered, holding for all $t$ as expected.

To compute the R\'enyi entropies, it is important to realize that the width of the annulus is no longer given by the difference between the values of $f(z)$ at the endpoints of $C$. In fact if we define 
\be 
\label{calW global}
{\cal W}=f(\textrm{i}\tau+\infty)-f(\textrm{i}\tau+\epsilon)\sim\log\left(\frac\beta{\pi\epsilon}\cos(2\pi\tau/\beta)\right)-\frac{2\pi\textrm{i}\tau}\beta\,,
\ee
this is complex. The actual width $W$ is the real part $\frac12({\cal W}+\overline{\cal W})$.

Thus, continuing $\tau\to\textrm{i}t$, the R\'enyi entropies from (\ref{SA_n generic}) are
\be
\label{SAn time-dep global}
S_A^{(n)} = \frac{c}{12}\left( n +\frac{1}{n}\right) \,\log\left( \frac{\beta}{\pi \epsilon}\, \cosh(2\pi t/\beta) \right) 
+ g_a + g_b + \textrm{corrections}\,.
\ee
The entanglement entropy obtained from this formula in the limit of large $t$ is 
$S_A\sim (\pi c/3\beta) t$, which is in agreement with the results of Ref.\,\cite{cc-05-global quench}.
Let us recall that, when the subsystem $A$ is an interval of finite length $\ell$, we have two entangling points and therefore the coefficient of the linear growth of $S_A$ for $t<\ell/2$ is twice the value we find here.

It is worth writing $\mathcal{W}$ and $\overline{\mathcal{W}}$ after analytic continuation:
\bea
\mathcal{W} |_{\tau = \textrm{i}t} &=& 
\log\left( \frac{\beta}{\pi \epsilon}\; e^{2\pi t/\beta}  \cosh(2\pi t/\beta)\right) +O(\epsilon)\,,
\\
\rule{0pt}{.8cm}
\overline{\mathcal{W}} |_{\tau = \textrm{i}t} &=& 
\log\left( \frac{\beta}{\pi \epsilon}\; e^{-2\pi t/\beta}  \cosh(2\pi t/\beta)\right) +O(\epsilon)\,.
\eea
As anticipated in Sec.\,\ref{sec:intro}, from these formulas it becomes clear that the large $t$ behavior of the R\'enyi entropies comes almost entirely from $\mathcal{W} |_{\tau = \textrm{i}t}$, that is, the right-movers.

\begin{figure}
\vspace{.3cm}
\hspace{.4cm}
\includegraphics[width=.95\textwidth]{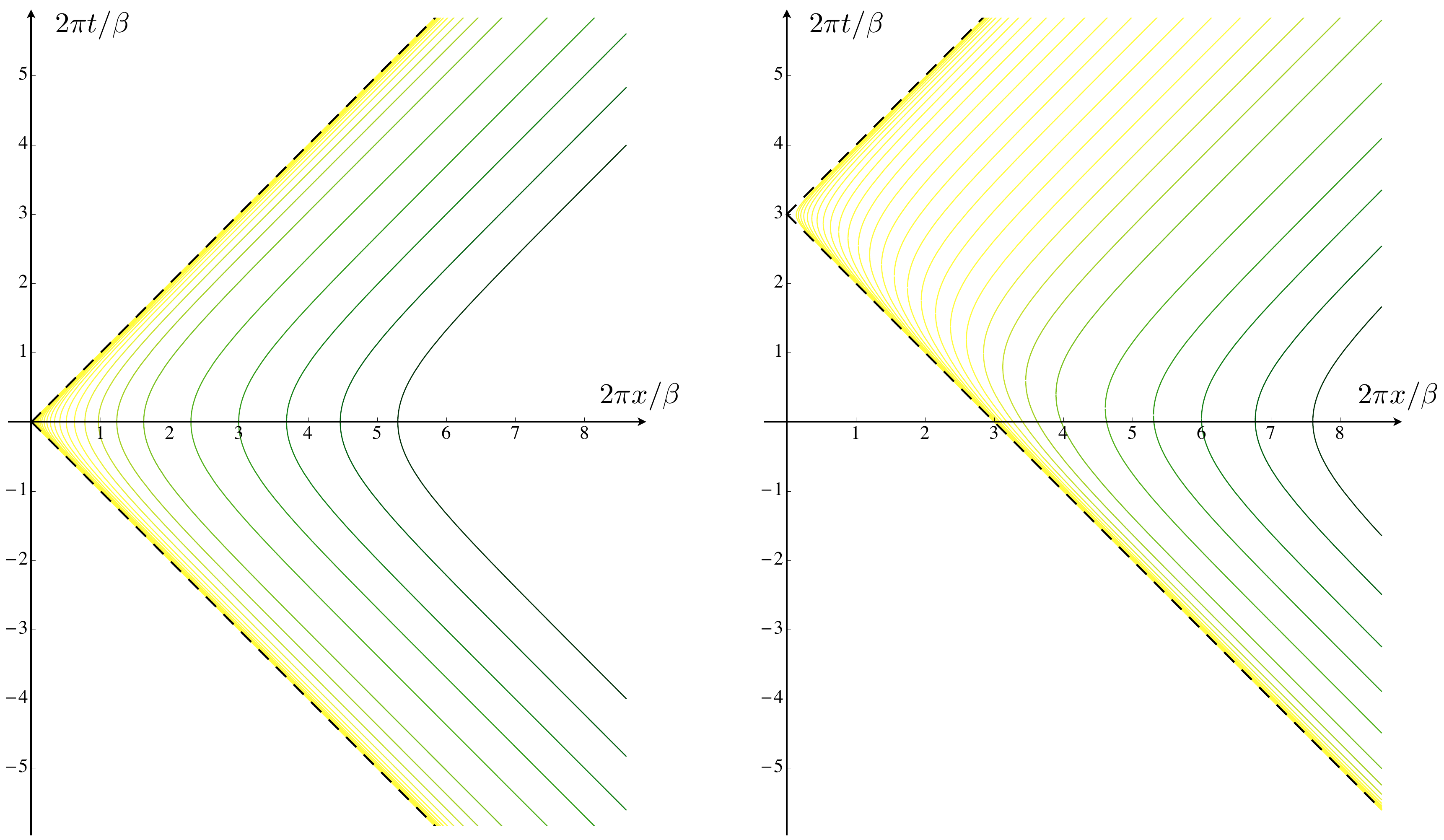}
\vspace{.2cm}
\caption{
Flows in Minkowski space given by (\ref{global Mflows}), corresponding to the modular hamiltonian 
(\ref{K_A global time v2}) at time $t_0$ after a global quench. 
Each curve is characterized by the value of the constant in the r.h.s. of (\ref{global Mflows}) and has the same color in both the panels. 
The flows are always contained in the Rindler wedge $x>|t-t_0|$, delimited by the dashed lines. 
At $t_0=0$ (left panel) these are hyperbolic only for $x\ll\beta$, corresponding to thermal behavior: for larger values they asymptote exponentially fast to $x=|t|$, corresponding to short-range entanglement. For $t_0\gg\beta$ the behavior is qualitatively different for $x<t_0$ and $x>t_0$. The asymmetry about $t=t_0$ is due to the different weights for R- and L-movers in (\ref{K_A global time v2}).
In the right panel $2\pi t_0 /\beta = 3$ and these features can be observed. 
}
\label{fig:global Mflows}
\end{figure}

It interesting to examine the flows in Minkowski space given by ${\rm Re}\,f(z)=\textrm{constant}$, being $f$ the expression in (\ref{fglobal}).
Making the analytic continuation given by $\tau\to \textrm{i}t_0$, $z \to x-t$ and $\bar{z} \to x+t$ (we now denote the observation time by $t_0$ to distinguish it from the Minkowski coordinate $t$), the equation describing these flows become
\be\label{global Mflows}
\frac{\sinh(\pi(t-x-t_0)/\beta)}{\cosh(\pi(t-x+t_0)/\beta)}\;
\frac{\sinh (\pi (t+x-t_0)/\beta)}{\cosh(\pi (t+x+t_0)/\beta)}
\,=\,\textrm{constant}\,.
\ee
These flows are illustrated in Fig.~\ref{fig:global Mflows}.

\subsection{Local quench}
\label{sec:local quench}

In this section we consider the local quench setup in a CFT in which two semi-infinite systems on the intervals $(-\infty,0)$ and
$(0,\infty)$, each with the same conformal boundary condition at $x=0$, are prepared in their respective ground states $|0\rangle_{L,R}$, and are then joined at $t=0$. The evolution hamiltonian is the translational invariant $H_{\textrm{\tiny CFT}}$ but the initial state is not translational invariant.

We elaborate on the analysis of this problem done in \cite{cc-07-local quench}.
 We in fact consider the state $e^{-\lambda H_{\textrm{\tiny CFT}}}\big(|0\rangle_L\otimes|0\rangle_R\big)$, with $\lambda>0$ playing a similar role to that of $\beta$ in the global quench since otherwise the energy is infinite. However, note that this state has finite, non-extensive energy above the ground state of $H_{\textrm{\tiny CFT}}$.

\begin{figure}
\vspace{.2cm}
\hspace{-.2cm}
\includegraphics[width=1.\textwidth]{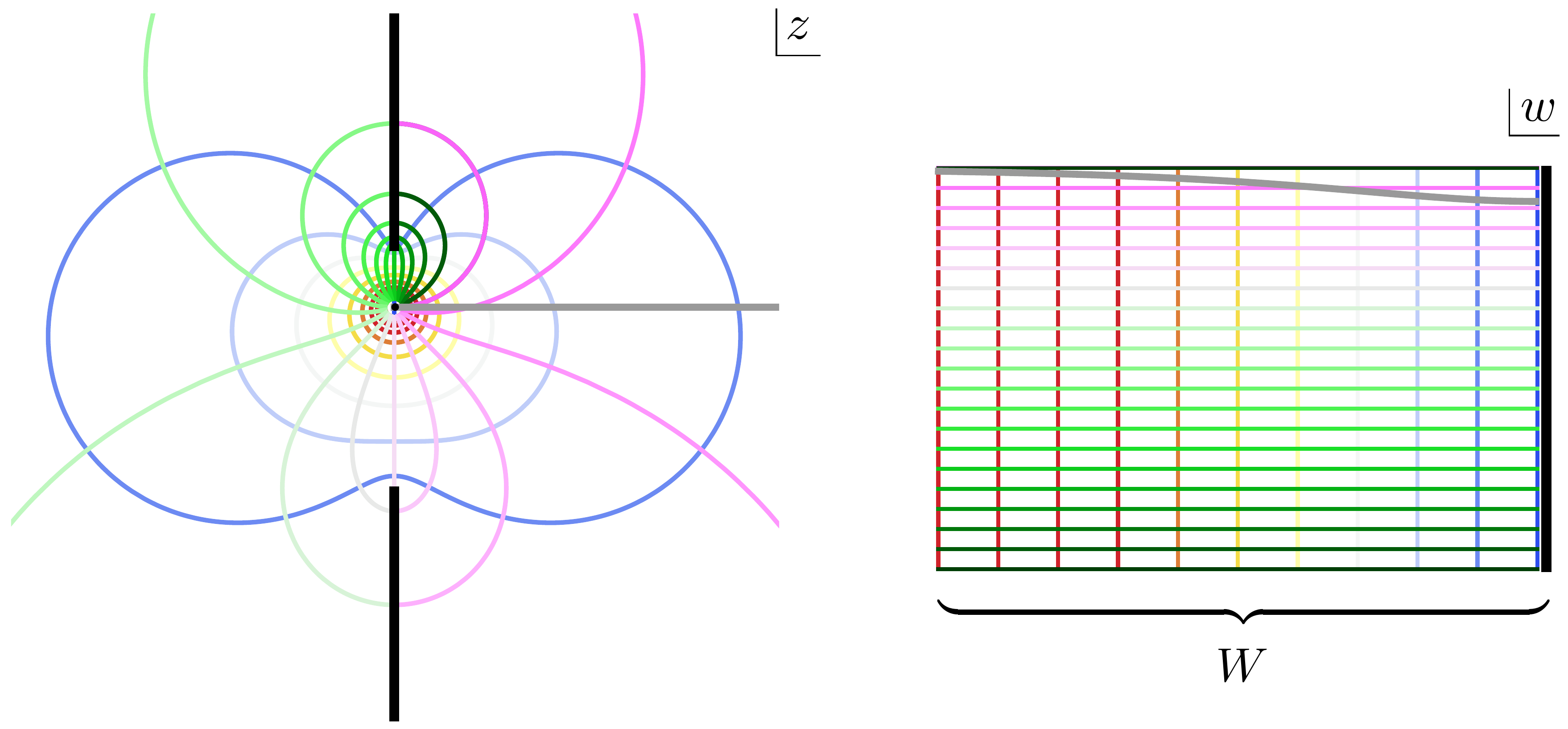}
\vspace{.1cm}
\caption{
Euclidean space-times characterising the case of the semi-infinite line after a local quench whose initial endpoint coincides with the defect, as discussed in Sec.\,\ref{sec:local quench}.
The conformal transformation (\ref{fmaplocal}) maps the space-time in the left panel onto the annulus in the right panel. 
The grey line in the left panel is $C = \{\textrm{i}\tau + x, \, x\geqslant 0\}$ is mapped onto the grey line in the right panel,
whose endpoints give the width of the annulus $W = \textrm{Re}(\mathcal{W})$, which provides the R\'enyi entropies (\ref{renyi local}).
} 
\label{fig:local}
\end{figure}

Following \cite{cc-07-local quench}, the euclidean space setup is as follows. 
Consider a plane with two slits along the imaginary axis parameterized by the complex coordinate $z$: one slit goes from $-\textrm{i} \lambda$ to $-\textrm{i} \infty$ and the other one from $\textrm{i} \lambda$ to $+\textrm{i} \infty$, being $\lambda >0$. Although we may consider various ways of bipartitioning the system, the simplest is to take $A=(0,\infty)$ and $B=(-\infty,0)$. 
Another possibility is studied in the Appendix \ref{app: local dec}. 
According to \cite{cc-07-local quench}, in the above plane we have to consider the semi-infinite line $C = \{z=\textrm{i}\tau + x, x\geqslant 0\}$, where $0<\tau<\lambda$.
This domain is shown in the left panel of Fig.\,\ref{fig:local}, where $C$ is given by the grey line. 
Let us remark that also in this case $C$ and $\overline{C}$ do not coincide.

Removing as usual a small disc around $z_0=\textrm{i}\tau$, the remainder may be mapped to the standard presentation of the annulus by
\be
\label{fmaplocal}
w=f(z) 
\,=\, 
\log\left(\frac{\sqrt{(\lambda^2 - \tau^2)(z^2+\lambda^2)} - \textrm{i} \tau \, z -\lambda^2}{\lambda(z-\textrm{i} \tau)} \right).
\ee
This may be found by first mapping to the right half plane, and then to the annulus (technical details on the construction of the conformal map (\ref{fmaplocal}) are given in  Appendix~\ref{app: maps}). 
Once again, it is a complex electrostatic potential, that due to a unit charge at $z_0=\textrm{i}\tau$ in the presence of conductors along $(x=0,|y|>\lambda)$.
The derivative of this conformal map is simpler, and reads
\be
\label{fprime local}
f'(z) = \frac{\sqrt{\lambda^2-\tau^2}}{(z-\textrm{i} \tau)\,\sqrt{z^2+\lambda^2}}\,.
\ee
Then, the entanglement hamiltonian (\ref{KAgeneral T barT}) in this case becomes 
\be
\label{K_A local 0}
K_A =
\int_0^{\infty} 
\frac{x\,\sqrt{(x+\textrm{i}\tau)^2+\lambda^2}}{\sqrt{\lambda^2-\tau^2}}\;
T(x+\textrm{i}\tau)\, dx
+
 \int_0^{\infty} 
\frac{x\,\sqrt{(x-\textrm{i}\tau)^2+\lambda^2}}{\sqrt{\lambda^2-\tau^2}}\;
\overline{T}(x-\textrm{i}\tau)\, dx\,.
\ee
Similarly to the global quench case, the integration domains of the two integrals in (\ref{K_A local 0}) are the same but the functions multiplying $T$ and $\overline{T}$ in the integrands are different.

Making the analytic continuation $\tau \to \textrm{i}t$, the entanglement hamiltonian (\ref{K_A local 0}) becomes
\be
\label{K_A local t}
K_A =
\int_0^{\infty} 
\frac{x\,\sqrt{(x-t)^2+\lambda^2}}{\sqrt{\lambda^2+t^2}}\;
T(x-t)\, dx
+
 \int_0^{\infty} 
\frac{x\,\sqrt{(x+t)^2+\lambda^2}}{\sqrt{\lambda^2+t^2}}\;
\overline{T}(x+t)\, dx\,.
\ee
As in the global case, $T(x-t)$ and $\overline T(x+t)$ here should be thought of as $T(x,t)$ and $\overline T(x,t)$ respectively.
Alternatively we may shift the $x$-integration variables to find 
\be
\label{K_A local t v2}
K_A =
\int_{-t}^{\infty} 
\frac{(x+t)\,\sqrt{x^2+\lambda^2}}{\sqrt{\lambda^2+t^2}}\;
T(x)\, dx
+
 \int_{t}^{\infty} 
\frac{(x-t)\,\sqrt{x^2+\lambda^2}}{\sqrt{\lambda^2+t^2}}\;
\overline{T}(x)\, dx\,,
\ee
where now the components of the stress tensor should be thought of as taken at $t=0$. 

The main features of the result (\ref{K_A local t}) or, equivalently (\ref{K_A local t v2}), have been already  anticipated in Sec.~\ref{sec:intro}.
It is worth analysing some special regimes of these formulas. 

When $t=0$ the entanglement hamiltonian reads
\be
K_A\big|_{t=0} = \int_{0}^{\infty} x\, \sqrt{(x/\lambda)^2+1}\; T_{00}(x)\, dx\,.
\ee
This agrees with the result (\ref{halfspace}) for the half-space for $x\ll\lambda$, but otherwise shows that the entanglement between the right and the left halves is short-ranged. 
For large $t$ the leading term of (\ref{K_A local t v2}) is given by Eq.\,(\ref{KA local large t}), which is also the expression obtained when $\lambda=0$.

Taking the limit $\lambda \to \infty$ of (\ref{K_A local t}) we get
\be
\lim_{\lambda \to \infty} K_A  =  \int_{0}^{\infty} x\, \big( T(x-t) +  \overline{T}(x+t)\big)\, dx
=\int_0^\infty x\,T_{00}(x,t)\,dx\,,
\ee
which is  (\ref{halfspace}) specialized to $d=2$, as it should, since now the whole system is in the ground state of $H$ at $t=0$.

In evaluating the R\'enyi entropies we again find that the width $W$ of the annulus is not simply given by the difference of the values of $f(z)$ at the endpoints of $C$, since this is complex. In fact, introducing this complex width as in the previous section, after the analytic continuation we find
\bea
\label{cal W local}
\mathcal{W} |_{\tau = \textrm{i}t} &=& 
\log\left( \frac{t^2+\lambda^2}{(\sqrt{t^2+\lambda^2} -t) \,\epsilon/2} \right) +O(\epsilon)\sim 3\log t 
\hspace{.6cm} \textrm{as}  \,\,t\to\infty\,,
\label{3logt}\\
\rule{0pt}{.8cm}
\label{cal bW local}
\overline{\mathcal{W}} |_{\tau = \textrm{i}t} &=& 
\log\left( \frac{t^2+\lambda^2}{(\sqrt{t^2+\lambda^2} + t) \,\epsilon/2} \right) +O(\epsilon))\sim \log t
\hspace{.72cm} \textrm{as}\,\,t\to\infty\,.
\label{logt}
\eea
The R\'enyi entropies can be found from (\ref{SA_n generic}), using $W=\frac12({\cal W}+\overline{\cal W})$:
\be
\label{renyi local}
S_A^{(n)} = 
\frac{c}{12}\left( n +\frac{1}{n}\right) \,
\log\left( \frac{t^2+\lambda^2}{\epsilon\, \lambda/2} \right) 
+ g_a + g_b + \textrm{corrections}\,.
\ee
For large $t$ we find $S^{(n)}_A = \tfrac{c}{6}(n+1/n) \log t+\dots$, which leads to $S_A = (c/3) \log t+\dots$ for the entanglement entropy, in agreement with the result of Ref.\,\cite{cc-07-local quench}.

However, we see from (\ref{3logt}) and (\ref{logt}) that in this case the R-movers contribute a factor 3 more than the L-movers to $S_A$, as discussed in Sec.\,\ref{sec:intro}.

In the Appendix\,\ref{app: local dec} we slightly generalize this setup by considering the case of a semi-infinite line $A$ after a local quench where the endpoint of $A$ and the point where the two semi-infinite lines are joined at $t=0$ do not coincide.

\subsection{Inhomogeneous quench}

In this section we compute the entanglement hamiltonian for a quench from an inhomogeneous state of the type first discussed in Ref.~\cite{sot-jc-08}. This has essentially the same form $e^{-(\beta/4)H}|b\rangle$ as that used for a translationally invariant
global quench in Sec.\,\ref{sec:global quench}, except that $\beta$ is taken to be slowly varying function of position $x$. 

Thus the euclidean path integral is now over the region $|{\rm Im}\,z|\leqslant \frac14\beta(x)$. We assume that $\beta(x)\to\beta_\pm$ as $x\to\pm\infty$, although it may be that $\beta_-\not=\beta_+$. This geometry may always be mapped to a uniform strip
$|{\rm Im}\,\zeta|\leqslant \frac14\beta$ by a conformal mapping $z\to\zeta=g(z)$. 
The mapping to the annulus is then $z\to w=f[g(z)]$ where
$f$ is the mapping in (\ref{fglobal}). 

From (\ref{KAgeneral T barT}) the entanglement hamiltonian is then
\be
K_A=\int_C\frac{T(z)}{f'[g(z)]g'(z)}\, dz+\int_{\overline C}\frac{\overline T(\bar z)}{\overline{f'[g(z)]g'(z)}}\, d\bar z\,.
\ee

Although the precise form of $g$ is complicated, it was pointed out in  \cite{sot-jc-08} that, as long as the support of $\beta'(x)$ is bounded, say within $|x|<l$, only the asymptotic behaviors of $g(z)$ as $z\to\pm\infty$ are relevant to the late time, large distance behavior of correlation functions and the entanglement entropy. Since in these limits $g$ is locally a scale transformation and real translation, we must have 
\begin{eqnarray}
g(z)&\sim&(\beta/\beta_+)(z+a_+)\,,\quad \hspace{.6cm} \textrm{as}  \,\, z\to+\infty\,,\\
      &\sim&(\beta/\beta_-)(z-a_-)\,,\quad \hspace{.6cm} \textrm{as} \,\, z\to-\infty \,,
\end{eqnarray}
where
\be
a_+=\lim_{x\to\infty}\left(\frac1{\beta_+}\int_0^x\beta(x')dx'-x\right)=O(l)\,,
\ee
and similarly for $a_-$. Note that then $g(0)=O(l)$. The cross-over region between these two asymptotic expressions is concentrated around $|x|=O(l)$. After analytic continuation $\tau\to\textrm{i}t$, only real values of $z$ and $g(z)$ enter the expressions.

Now suppose that $A=(x_0,\infty)$ and $B=(-\infty,x_0)$, where for simplicity we take $x_0\gg\xi$.
In the same approximation that leads to (\ref{Kglobalapprox}) we may then write
\be
K_A=\frac\beta\pi\int_{g(x_0-t)<g(x)<g(x_0+t)}\frac{T(x)}{g'(x)} \,dx\,.
\ee
Then the result depends on whether $x<0$ is in the past light cone of $A$. If $x_0>t+O(l)$, we have simply
\be 
K_A\sim\frac{\beta_+}\pi\int_{x_0-t}^{x_0+t}T(x) \, dx\,.
\ee
However, if $t> x_0+O(l)$ we have
\be
K_A\sim\frac{\beta_-}\pi\int_{x_0-t-a_-}^{O(l)}T(x) \, dx+\frac{\beta_+}\pi\int_{O(\xi)}^{x_0+t}T(x) \,dx\,,
\ee
where the contribution from $x=O(\xi)$ depends on the details of $\beta'(x)$. This has an obvious quasiparticle interpretation:
the entanglement between $A$ and $B$ is again due only to R-movers. Those that originate from $x<0$ have effective inverse temperature $\beta_-$, while those from $x>0$ have inverse temperature $\beta_+$. Note, however the shift in the lower limit
in the first term: the conformal map $g$ distorts the light-cones near $x=O(\xi)$, so that the quasiparticles coming from $x<0$ are red or blue-shifted as they enter the region $x>0$. 

We stress that we have discussed only the simplest features of an inhomogeneous quench. It would be possible to obtain much more explicit expressions for particular choices of $\beta(x)$.

\section{Discussion}
\label{sec5}

In this paper we have discussed examples in 2d CFT where the entanglement (modular) hamiltonian $K_A$ may be written as an integral over the energy-momentum tensor times a local weight. A sufficient condition is that the euclidean space-time region describing the traces of powers of the reduced density matrix (once small slits or discs around the entangling points have been removed) is topologically an annulus. In these cases, a conformal mapping shows that the universal part of the spectrum of $K_A$ is always that of a boundary CFT with appropriate boundary conditions, in agreement with \cite{tachikawa-14}, and suggested by numerical studies in \cite{lauchli}. One may ask whether this condition is also necessary. The answer appears to be positive.
If we denote the intersection of $A$ with a constant time slice by $C$ as before, then the space-time generated by $e^{-\theta K_A}$ with $0\leqslant \theta<2\pi$ is topologically $C\times S_1$. If $C$ has a single component this is an annulus, but if $C$ has more than one component, this is a product of annuli, which has a different genus from the euclidean space-time describing $\rho_A$.

\begin{figure}
\vspace{.5cm}
\includegraphics[width=1.0\textwidth]{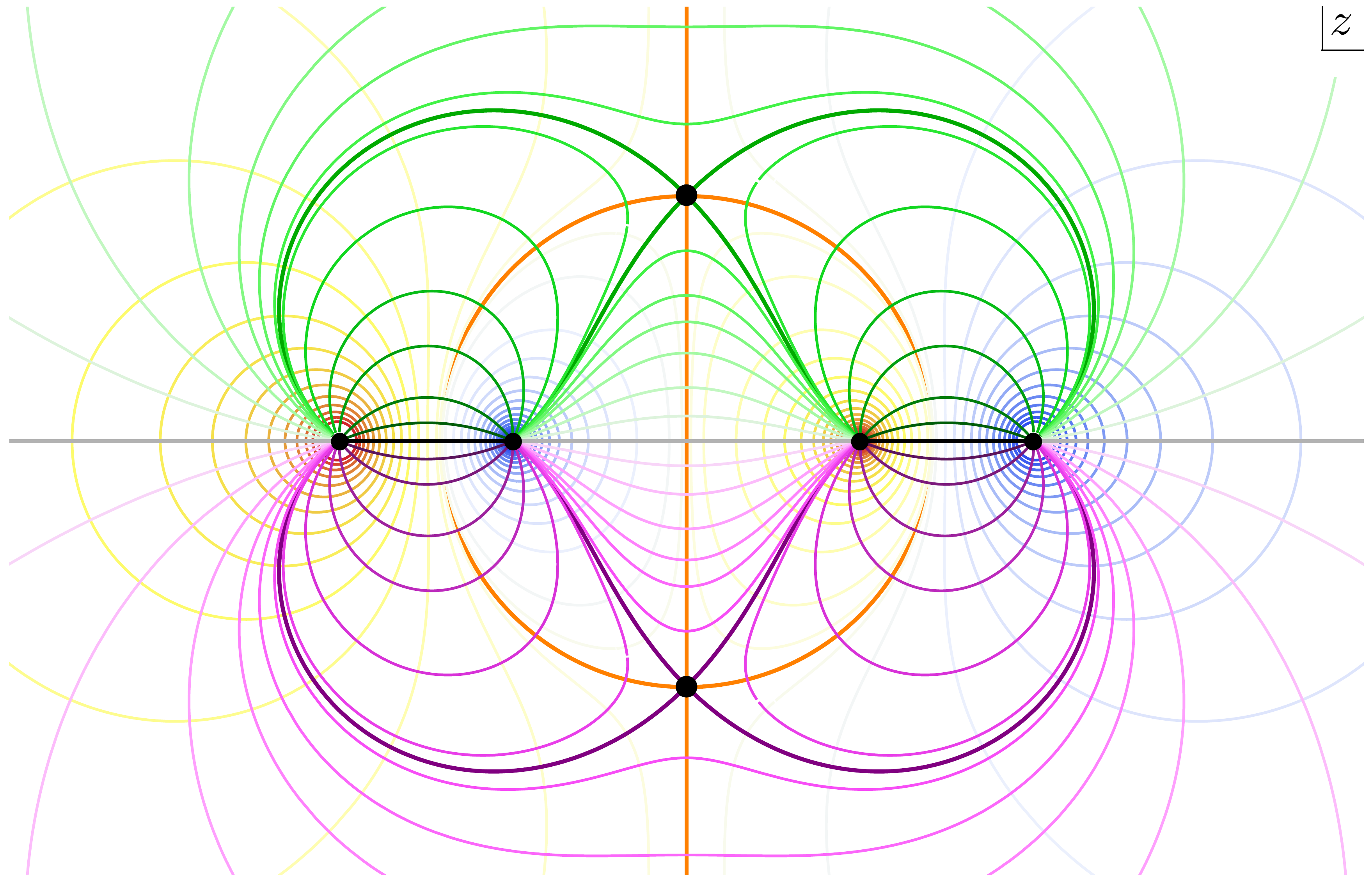}
\vspace{-.1cm}
\caption{
The action of $K_A$ as defined by Eqs.\,(\ref{KAgeneral}) and (\ref{f2intervals}) when $A$ consists of two disjoint intervals whose endpoints are the black dots on the horizontal axis. The evolution under $v={\rm Im}\,f$ propagates $A$ into the upper half plane, but it stops when it hits the point $\textrm{i}\sqrt{ab}$ (black dot in the upper half plane) where $f'$ vanishes. The remainder of the euclidean space-time region is covered by a similar propagation with $K_B$, but these have to be joined by operators $(X,X^{\dag})$ which effectively sew together the states on the boundary between these two regions.
} 
\label{fig:2int}
\end{figure}

It is interesting to examine what goes wrong if we naively apply our methods to the case when $A$ is the union of two disjoint 
intervals, and $B$ is the complement in the infinite line. Without loss of  generality we may take these to be $A_1=(-b,-a)$ and $A_2=(a,b)$. This problem has been studied exhaustively in the literature \cite{2int-refs, ch-2int-KA-ff,casini-2int,longo-disjoint}. Using the electrostatic analogy, we might try to use the mapping function
\be\label{f2intervals}
w=u+\textrm{i} v=f(z)=\log\left(\frac{(z+b)(z-a)}{(z+a)(z-b)}\right) ,
\ee
and to use (\ref{KAgeneral}) as the form for $K_A$. Unfortunately this fails, because, as suggested by the above argument,
$e^{-\theta K_A}$ does not cover the euclidean space-time smoothly as $\theta$ increases from zero to $2\pi$. 
The curves $v=\theta$ are, for small enough $\theta$,
homotopic to $C$, but at a particular value of $\theta=\theta_0$ (depending on the cross-ratio of the four points), the evolution becomes singular. This is related to the fact that $f'(\pm\textrm{i}\sqrt{ab})=0$, so the mapping fails to be conformal there. So all we can do with this $K_A$ is cover part of the space-time (see Fig.~\ref{fig:2int}).

The remainder can be covered by a similar expression $K_B$ which is an integral over $B$. $K_A$ and $K_B$ are operators in ${\cal H}_A$ and ${\cal H}_B$ respectively. The sewing together of these two evolutions involves an operator $X$ which is a map from ${\cal H}_A\to {\cal H}_B$. Thus we may write
\be
\rho_A =e^{-\theta_0K_A}\,X^{\dagger}\,e^{-(2\pi-2\theta_0)K_B}\,X\,e^{-\theta_0K_A}\,,
\ee
but there is no sense in which we may write $-\log\rho_A$ as an integral over $A$ or $B$ alone. The existence of this singularity in the imaginary time evolution implies the non-validity of the KMS condition for this candidate for the modular hamiltonian, since this assumes that the real time evolution may be analytically continued to imaginary time with no obstructions.

It is interesting, therefore, that for a free chiral fermion Casini and Huerta \cite{casini-2int} have managed to circumvent this difficulty. The flow they use is identical to that along constant ${\rm Re}\,f(z)$ as given by (\ref{f2intervals}), but in addition it mixes the fermion field at $z$ with that at $z'$, where $f(z')=f(z)$ (that is, $zz'=-ab$.) 
This mixing appears to allow the smooth passage through the singularities. Indeed, in Ref.~\cite{longo-disjoint} it is shown explicitly that the Casini-Huerta flow satisfies the KMS condition (and the result is generalized to an arbitrary number of intervals). It remains to be seen whether this kind of orbifold construction works for a general CFT.

We remark that a similar difficulty arises in the case when $A$ is a single sub-interval of a circle, at finite
temperature (so that the euclidean space-time is a torus.) In that case also, the simplest candidate for $f$ is non-conformal at 2 points: since $f'(z)$ has 2 poles on the torus it must also have 2 zeroes.

We have also emphasized that the procedure in (\ref{SA?}) by which the entanglement entropy is computed as an integral over the thermal entropy density coming from a position-dependent effective temperature, only gives the correct result by virtue of special properties of 2d CFT, namely that it may be conformally mapped to an annulus which corresponds to a uniform temperature on an interval of a certain width. There is no reason why this procedure should work more generally, except perhaps in special limits \cite{swingle-16}. One such limit is as the R\'enyi index $n\to0^+$ in higher dimensions. Consider the simplest case when $A$ is the half-space $x_1>0$, so (\ref{halfspace}) applies. The euclidean space-time region for the R\'enyi entropies is then a wedge-shaped region $0\leqslant \theta<2\pi n$ where $\theta=\tan^{-1}(x_0/x_1)$ in cylindrical coordinates, with $\theta=0$ and $\theta = 2\pi n$ identified. In that case, from (\ref{halfspace}), the effective inverse temperature is $\beta_{\rm eff}\propto nx_1$. As long as 
$\beta_{\rm eff}^{-1}\partial_{x_1}\beta_{\rm eff}\sim n\ll1$, it should be possible to approximate $\log{\rm Tr}\rho_A^n$ by an integral over the appropriate free energy density for a uniform temperature. For a $d$-dimensional CFT, this behaves
$\propto \sigma/\beta^{d-1}$, where $\sigma$ is the appropriate generalization of the Stefan-Boltzman constant. We therefore find a term 
\be
S_A^{(n)}\propto\frac\sigma{n^{d}}\int_\epsilon^\infty \frac{dx_1}{x_1^{d-1}}\,\int d^{d-2}x_\perp\,,
\ee
as $n\to0^+$. This gives the correct logarithmic dependence on the cut-off for $d=2$. For $d>2$ the integration over the remaining $d-2$ coordinates $x_\perp$ gives a factor of the area of the boundary $\partial A$, and this is multiplied by a UV divergent factor $1/\epsilon^{d-2}$, as expected. 
So, although this argument is probably correct for $n\ll 1$, it applies only to the non-universal area law term for $d>2$.

\section*{Acknowledgements}
We thank M.~Mezei for making a crucial comment on an earlier version of our results, and also P.~Calabrese, H.~Casini, H.~Katsura, A.~L\"auchli and B.~Swingle for useful comments and references.

The authors are grateful to the KITP, Santa Barbara and the YITP, Kyoto, and ET also to UC Berkeley, for hospitality and support while part of this work was carried out. 
ET has been supported by the ERC under Starting Grant  279391 EDEQS.

\begin{appendices}

\section*{Appendices}

\section{On the conformal maps for the cases with an external boundary}
\label{app: maps}

In this appendix we discuss the construction of some conformal maps $w=f(z)$ sending the space-time characterising the physical problem (parameterized by the complex coordinate $z$) onto the annulus (described by the complex coordinate $w=u+\textrm{i} v$), which allow to write the corresponding entanglement hamiltonians by employing Eq.\,(\ref{KAgeneral T barT}).
We focus on those cases where, in the $z$-spacetime, the internal boundary is the small disc encircling the branch point $z_0$, while the other boundary is an external one along which the same boundary condition is imposed. 
The annular domain here is the one introduced in Sec.\,\ref{subsec:bdy} (see the right panel of Fig.\,\ref{fig:interval semi-infinite line T=0}).
Explicit examples of these maps have been used in Sec.\,\ref{subsec:bdy}, \ref{sec:2bdy}, \ref{sec:global quench} and \ref{sec:local quench}.

As first step to construct $f(z)$, we have to find the conformal map $\xi=\xi(z)$ which sends the $z$-spacetime onto the right half plane (here parameterized by the complex coordinate $\xi$), namely $\textrm{Re}(\xi) \geqslant 0$.
The external boundary in the $z$-spacetime is mapped onto the vertical line $\textrm{Re}(\xi)=0$, while the small disc around the branch point is sent onto a small disc around $\xi_0 = \xi(z_0)$.

Assuming that the map $\xi(z)$ is known, let us construct the conformal transformation $w(\xi)$ which maps the right half plane onto the annulus,
requiring also that a small disc around the point $\xi_0$ is sent onto the boundary $\lim_{u\to -\infty} (u+\textrm{i} v)$ of the annulus.
This can be done by introducing the unit disc (parameterized by the complex coordinate $\zeta$), i.e. the domain $|\zeta| \leqslant 1$, as intermediate step. 
Indeed, the conformal transformation $w=\log(\zeta)$ maps an infinitesimal disc around the origin $\zeta =0$ and the external boundary $|\zeta|=1$ onto the boundaries $\lim_{u\to -\infty} (u+\textrm{i} v)$ and $\textrm{Re}(w)=0$ of the annulus respectively (see the middle and right panels of Fig.\,\ref{fig:interval semi-infinite line T=0}).  
Thus, $w(\xi) = w(\zeta(\xi))$, where $\zeta(\xi)$ is  the conformal map which sends the right half plane onto the unit disc and such that $\zeta(\xi_0)=0$.
The latter transformation can be found by composing $\xi \to \tilde{\zeta} = (\xi-1)/(\xi+1)$, which sends the right half plane onto the unit disc, with a generic $SU(1,1)$ transformation, i.e. $\tilde{\zeta} \to \zeta =  \tfrac{\alpha\,\tilde{\zeta} + \gamma}{\bar{\gamma}\,\tilde{\zeta} + \bar{\alpha}} = \tfrac{\alpha}{\bar{\alpha}}\, \tfrac{\tilde{\zeta} + \eta}{\bar{\eta}\,\tilde{\zeta} + 1}$, where $\eta = \gamma/\alpha$, mapping the unit disc onto itself. 
The rotational symmetry about the origin $\zeta =0$ allows us to set $\alpha>0$.
Then, the requirement that $\zeta(\xi_0)=0$ leads to $\eta = (1-\xi_0)/(1+\xi_0)$.

The final result for the conformal map sending the initial space-time with one external boundary to the annulus is given by 
\be
\label{zeta map general 0}
f(z)
= \log\left[
\left(\frac{1+\bar{\xi}_0}{1+\xi_0} \right)
\frac{\xi(z)-\xi_0}{\xi(z)+\bar{\xi}_0} \, \right] ,
\ee
where the conformal transformation $\xi(z)$ and the point $\xi_0$ are determined by the initial physical situation. 
Taking the derivative of (\ref{zeta map general 0}), one finds  
\be
\label{fprime general 0}
f'(z) 
= \frac{2\textrm{Re}(\xi_0)\, \, \xi'(z)}{\big(\xi(z)-\xi_0\big)\big(\xi(z)+\bar{\xi}_0\big)} \,,
\ee
which enters in the entanglement hamiltonian (\ref{KAgeneral T barT}).
Notice that in all the examples we consider  $\xi'(z)= \xi(z)/p(z)$, where $p(z)$ is a simple function.

In all the static examples addressed in the main text, the function $f(z)$ entering in the integrands of (\ref{KAgeneral T barT})  is real and, as for the integrations domains $C$ and $\overline{C}$, we have that $C= \overline{C}=A$. 
Thus, the entanglement hamiltonian (\ref{KAgeneral T barT}) becomes a single integral over $A$ of $T_{00}$ multiplied by the proper real function, as already remarked in Sec.\,\ref{sec:intro}.

In the expression (\ref{KAgeneral T barT}) for the entanglement hamiltonian, the integration domain $C$ is a line in the $z$-domain which starts at $z_0+\epsilon$ and ends at a point $z_b$ on the external boundary.
The endpoints of $C$ and the conformal map (\ref{zeta map general 0}) give us the R\'enyi entropies.
Indeed, let us introduce the following complex quantity 
\be
\label{calW def}
\mathcal{W}
=
f(z_b) - f(z_0 +\epsilon)
=
\log \left(\frac{\xi_0 +\bar{\xi}_0}{\epsilon\, \xi'(z_0)} \right)
+ \log \left(\frac{\xi(z_b)-\xi_0}{\xi(z_b)+\bar{\xi}_0} \right) + O(\epsilon)\,.
\ee
In the generic situation, $\textrm{Im}[f(z_b) ] \neq \textrm{Im}[f(z_0 +\epsilon) ]$ and  the width of the annular region is $W = \textrm{Re}[f(z_b)] - \textrm{Re}[f(z_0 +\epsilon)] = \textrm{Re}(\mathcal{W})$, which is given by
\be
\label{W + barW}
W=
\frac{\mathcal{W} + \overline{\mathcal{W}}}{2} \,=\,
\log (q_0/\epsilon)
+ \log \left|\frac{\xi(z_b)-\xi_0}{\xi(z_b)+\bar{\xi}_0} \right| + O(\epsilon)\,,
\qquad
q_0^2 
\equiv
\frac{(\xi_0+\bar{\xi}_0)^2}{|\xi'(z_0)|^2}\,.
\ee
The R\'enyi entropies are obtained by plugging (\ref{W + barW}) into  (\ref{SA_n generic}).

When $\textrm{Im}[f(z_b) ] = \textrm{Im}[f(z_0 +\epsilon) ]$ we have that $\mathcal{W}$ is real and equal to the width $W$ of the annulus.
This is always the case for the static examples that we have addressed.

The simplest case included in the above discussion is the static example considered in Sec.\,\ref{subsec:bdy}, where the initial space-time is the left half plane $\textrm{Re}(z) \leqslant 0$ (see the left panel of Fig.\,\ref{fig:interval semi-infinite line T=0}). 
Thus, $\xi(z) = -z$ and $\xi_0 = R$. 
The conformal map (\ref{zeta map general 0}) becomes (\ref{fmap single interval}) and its derivative (\ref{fprime general 0}) gives $f'(z) =2R/(R^2-z^2)$, which leads to the entanglement hamiltonian (\ref{KAball}) for $d=2$, as expected.

A less trivial static case is the one discussed in Sec.\,\ref{sec:2bdy}, where the initial space-time is $-L/2  \leqslant \textrm{Re}(z) \leqslant L/2$.
For this example the conformal map sending the initial space-time onto the right half plane is given by $\xi(z) = e^{\textrm{i}\pi z/L}$.
By specialising the argument of the logarithm in (\ref{zeta map general 0}) to this case, one gets the argument of the logarithm of (\ref{logsincos}) multiplied by the imaginary unit, which can be reabsorbed by employing the rotational invariance of the unit disc parameterized by $\zeta$.

Also the conformal mappings entering in the time-dependent examples are special cases of (\ref{zeta map general 0}).
In particular, for the global quench scenario considered in Sec.\,\ref{sec:global quench} the initial space-time is $-\beta/4  \leqslant \textrm{Im}(z) \leqslant \beta/4$ (Fig.\,\ref{fig:global}, left panel) and therefore $\xi(z) = e^{2\pi  z/\beta}$, which is a rotated version of corresponding map occurring  in the previous static example. 
Now $\xi_0 = e^{2\pi \textrm{i} \tau /\beta}$ and (\ref{zeta map general 0}) reduces to (\ref{fglobal}). 
The complex quantity $\mathcal{W}$ in (\ref{calW global}) can be found also by specialising (\ref{calW def}) to this case. 

As for the local quench discussed in Sec.\,\ref{sec:local quench}, the initial space-time is given in the left panel of Fig.\,\ref{fig:local} and the conformal transformation $\xi(z)=z/\lambda+\sqrt{(z/\lambda)^2+1}$ maps it onto the right half plane  \cite{cc-07-local quench}.
Then, since $\xi_0 = \sqrt{1-(\tau/\lambda)^2}+\textrm{i}\tau/\lambda$, which has $|\xi_0| =1$, we can specialize (\ref{zeta map general 0}) to this example, finding (\ref{fmaplocal}). 
Moreover, by specifying (\ref{calW def}) to this case, the expressions (\ref{cal W local}) and (\ref{cal bW local}) are obtained.

\section{Local quench: decentered case}
\label{app: local dec}

In this appendix we consider a local quench where, at $t=0$, two semi-infinite lines are joined at some point $x=0$ (the defect point).
Again, we are interested in the reduced density matrix of a semi-infinite line $A$ but, differently from Sec.\,\ref{sec:local quench}, now the entangling point $x_0$, which separates the semi-infinite lines $A$ and $B$, does not coincide with the defect point. 
Without loss of generality, we can choose $x_0 \geqslant 0$, and the special case $x_0 =0$ corresponds to the example discussed in Sec.\,\ref{sec:local quench}.

The initial euclidean space-time, parameterized by the complex coordinate $z$, is the one introduced in \cite{cc-07-local quench},
but in this case $z_0 = x_0 + \textrm{i}\tau$, with $0<\tau<\lambda$.
Thus, $C = \{ z=z_0 + x, \, x\geqslant 0\}$ and $\overline{C} = \{ z=x_0  - \textrm{i}\tau + x, \, x\geqslant 0\}$.
Like in the other time-dependent examples, $C$ and $\overline{C}$ do not coincide.

From the conformal map $\xi(z)$ given in Sec.\,\ref{sec:local quench}, which sends the $z$-plane onto the right half plane, in this case the image of the point $z_0$ is
\be
\label{xi_0 dec}
\xi_0 =
\frac{1}{\lambda}\Big(
\sqrt{x_0^2+ \lambda^2 -\tau^2 +   \textrm{i} \, 2 \tau x_0 }  +x_0 +\textrm{i} \, \tau
\Big) \, ,
\ee
and  $|\xi_0| > 1$ when $x_0 > 0$.
The map $w= f(z) $ to the annulus can be written by specifying (\ref{zeta map general 0}) to this case and the result is
\be
f(z) 
\,=\, 
\log\left[
\left(\frac{1+\bar{\xi}_0}{1+\xi_0} \right)
\frac{(\xi_0 + \bar{\xi}_0) \sqrt{(z/\lambda)^2+1} -(\xi_0 - \bar{\xi}_0)\, z/\lambda - |\xi_0|^2 - 1}{
2\bar{\xi}_0\, z/\lambda +\bar{\xi}_0^2 - 1}\, \right] ,
\ee
whose derivative reads
\be
\label{fprime local dec}
f'(z)  \,=\,
\frac{2\,\textrm{Re}(\xi_0) \, w(z)}{
\big\{ 2\,w(z)\big[ z/\lambda - \textrm{i}\, \textrm{Im}(\xi_0) \big] +1 -|\xi_0|^2\big\} \,
\sqrt{z^2+\lambda^2}}\,,
\ee
which can be found also by specialising (\ref{fprime general 0}) to this case.
Notice that for $x_0=0$ we have $|\xi_0|=1$ and a significative simplification occurs in (\ref{fprime local dec}), which becomes (\ref{fprime local}).

The expression (\ref{fprime local dec}) allows to write the entanglement hamiltonian by specifying (\ref{KAgeneral T barT}) to this case, but
the final formula is quite complicated and  we do not report it here.

As for the R\'enyi entropies (\ref{SA_n generic}), one can evaluate the width $W$ of the annulus by using (\ref{W + barW}).
In this case $z_b = \lim_{x\to +\infty}(x+\textrm{i}\tau)$ and $|\xi(z_b)| \to \infty$, therefore the second term in the r.h.s. of (\ref{W + barW}) vanishes.
Instead, as for the first term, here we find
\be
q_0^2 
=
\frac{\sqrt{(z_0^2 + \lambda^2)(\bar{z}_0^2 + \lambda^2)}\, 
\big( z_0 + \bar{z}_0 +  \sqrt{z_0^2 + \lambda^2} +  \sqrt{\bar{z}_0^2 + \lambda^2}  \,\big)^2
}{
\big( z_0 +  \sqrt{z_0^2 + \lambda^2} \,\big) \big( \bar{z}_0 +  \sqrt{\bar{z}_0^2 + \lambda^2} \, \big)}\,.
\ee

\begin{figure}
\vspace{.3cm}
\begin{center}
\includegraphics[width=.8\textwidth]{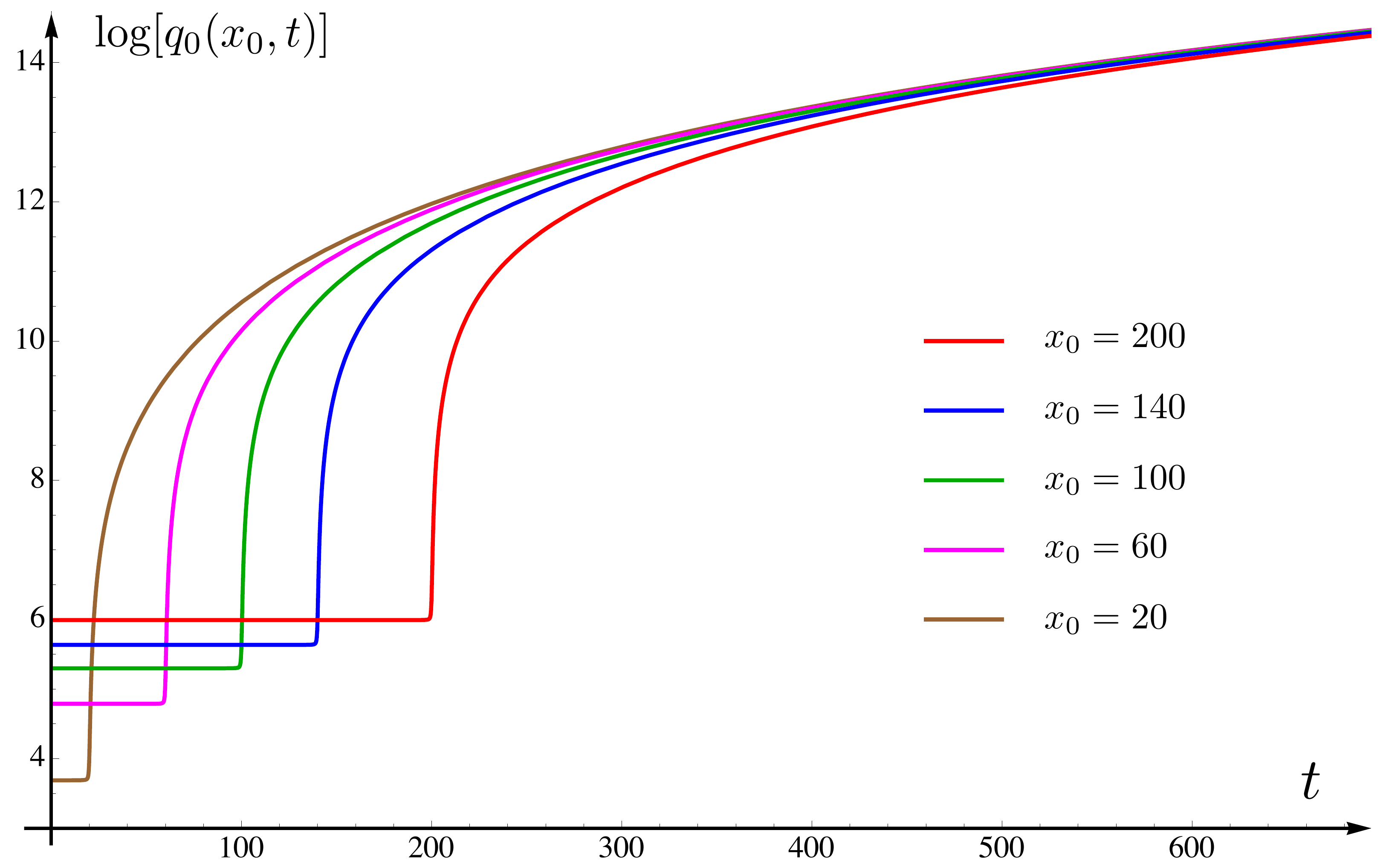}
\end{center}
\vspace{-.0cm}
\caption{Local quench in the decentered case discussed in Appendix\,\ref{app: local dec}: $\log[q_0(x_0,t)]$ as function of $t$ from Eq.\,(\ref{q_0 def local dec}) for some values of $x_0 \gg \lambda$ (here $\lambda=0.5$).
A sharp transition occurs at $t \simeq x_0$ between the constant value and the logarithmic behavior.
} 
\label{fig:local dec q0}
\end{figure} 

Making the analytic continuation $\tau \to \textrm{i}t$, we have that $z_0 \to x_0 - t$ and $\bar{z}_0 \to x_0 + t$. 
The above expression becomes $q_0^2 |_{\tau =\textrm{i}t} = q_0(x_0,t)^2$ given by
\be
\label{q_0 def local dec}
q_0^2 \big|_{\tau =\textrm{i}t}
=
\frac{\sqrt{((x_0-t)^2 + \lambda^2)((x_0+t)^2 + \lambda^2)}\, 
\big( 2x_0 +  \sqrt{(x_0-t)^2 + \lambda^2} +  \sqrt{(x_0+t)^2 + \lambda^2}  \,\big)^2
}{
\big( x_0-t +  \sqrt{(x_0-t)^2 + \lambda^2} \,\big) \big( x_0+t +  \sqrt{(x_0+t)^2 + \lambda^2} \, \big)}\,.
\ee
Notice that $q_0/\lambda$ is a function of $x_0/\lambda$ and $t/\lambda$ only. 
From (\ref{SA_n generic}), we have that the R\'enyi entropies read $S_A^{(n)} = \tfrac{c}{12} (n+1/n) \log[q_0(x_0,t)/\epsilon] +\dots$.

In Fig.\,\ref{fig:local dec q0} we show $\log[q_0(x_0, t)]$ as function of $t$ for some values of $x_0 \gg \lambda$. 
In this regime of parameters, the curve displays a sudden change at $t \simeq x_0$.
This sharp transition is smoothed out as $\lambda$ increases and becomes $\lambda \sim x_0$.
In the special case of $x_0=0$, the expression (\ref{q_0 def local dec}) simplifies to $q_0(0 , t) = 2(t^2+\lambda^2)/\lambda$, 
which is the result entering in (\ref{renyi local}), as expected.

Two interesting regimes to consider are given by small and large $t$.
In these limits (\ref{q_0 def local dec}) becomes respectively
\be
\label{q0 small large t}
q_0(x_0 , t)  = \left\{
\begin{array}{ll}
\displaystyle  2\,\sqrt{x_0^2+\lambda^2} + \frac{2\lambda^2 \, t^2}{(x_0^2+\lambda^2)^{3/2}} + O(t^4)
 \hspace{1cm} 
&   t\ll 1 \,,
\\
\rule{0pt}{.8cm}
\displaystyle  \frac{t^2}{\lambda/2} -\frac{x_0^2 - \lambda^2}{\lambda/2}+O(1/t^2)
&    t \gg 1\,.
\end{array}\right.
\ee
When $\lambda \sim \epsilon$ in (\ref{q0 small large t}), the result of \cite{cc-07-local quench} for this setup is recovered.

\end{appendices}

\end{document}